\documentstyle[12pt]{article}
\if@twoside  m
\else
\fi
\renewcommand{\theequation}{\arabic{section}.\arabic{equation}}

\newcommand{\ie}{{\em i.e.}}
\newcommand{\eg}{{\em e.g.}}
\newcommand{\cf}{{\em cf. }}
\newcommand{\etc}{{\em etc. }}
\newcommand{\viz}{{\em viz. }}
\newcommand{\QED}{\mbox{\rule[-1.5pt]{6pt}{10pt}}}
\newcommand{\lhs}{{\em lhs}}
\newcommand{\rhs}{{\em rhs }}
\newcommand{\tr}{{\rm tr\,}}
\newcommand{\re}{{\rm Re\,}}
\newcommand{\K}{I\!\!K}
\newcommand{\R}{I\!\!R}
\newcommand{\Z}{Z\!\!\!Z}
\newcommand{\HH}{{\cal H}}

\newcommand{\LL}{{\cal L}}
\newcommand{\MM}{{\cal M}}
\newcommand{\OO}{{\cal O}}
\newcommand{\SS}{{\cal S}}
\newcommand{\eps}{\varepsilon}
\newtheorem{claim}{Claim}[section]
\newtheorem{theorem}[claim]{Theorem}
\newtheorem{lemma}[claim]{Lemma}

\begin{document}
\title{A single--mode quantum transport in \\ serial--structure
geometric scatterers}
\date{15/9/2000}
\author{P.~Exner,$^{1,2}$ M.~Tater,$^1$ and D.~Van\v{e}k$^3$}
\maketitle
\begin{quote}
{\small \em 1 Nuclear Physics Institute, Academy of Sciences,
CZ--25068 \v Re\v z near Prague, \\ 2 Doppler Institute, Czech
Technical University, B\v rehov{\'a} 7, CZ-11519 Prague, \\ 3
Department of Mathematics, FNSPE, Czech Technical University, \\
\phantom{a)}Troja\-nova 13, CZ--12000 Prague, Czech Republic} \\
\rm
\phantom{e)}exner@ujf.cas.cz, tater@ujf.cas.cz, vanekd@Alenka.ufa.cas.cz

\end{quote}

\begin{abstract}
\noindent We study transport in quantum systems consisting of a
finite array of $\,N\,$ identical single--channel scatterers. A
general expression of the S matrix in terms of the
individual--element data obtained recently for potential
scattering is rederived in this wider context. It shows in
particular how the band spectrum of the infinite periodic system
arises in the limit $\,N\to\infty\,$. We illustrate the result on
two kinds of examples. The first are serial graphs obtained by
chaining loops or T--junctions. A detailed discussion is presented
for a finite--periodic ``comb"; we show how the resonance poles
can be computed within the Krein formula approach. Another example
concerns geometric scatterers where the individual element
consists of a surface with a pair of leads; we show that apart of
the resonances coming from the decoupled--surface eigenvalues such
scatterers exhibit the high--energy behavior typical for the
$\,\delta'$ interaction for the physically interesting couplings.
\end{abstract}

\tableofcontents


\section{Introduction}

A rapid progress in experimental solid state physics has expanded
dramatically the list of situations in which consequences of the
basic equations of quantum mechanics may be tested, since the
interaction is prescribed by the experimentalist by the shape design
of the structure in question, material choice, \etc One of the
frequently occuring cases in mesoscopic transport involves a passage
of a quantum particle through a {\em serial} --- or finitely periodic
--- structure obtained by arraying a certain number $\,N\,$ of
identical scatterers.

Our aim in the present paper is to study this situation under the
assumption that the individual scatterers have a single transport
mode. For a collection of mesoscopic elements connected by quantum
wires, this is certainly an idealization. We can adopt this
approximation provided the transverse modes in the wires are well
separated and the distances between the scatterers are large enough
so that the intermode coupling and influence of the evanescent modes
can be neglected.

Such a single--mode transport is often investigated in literature,
with the S--matrix obtained either ``inductively" by adding the
scatterers successively or by means of the transfer matrix method. It
is difficult to collect all relevant references but a representative
sample is given in \cite{ET}; an extension of this ``factorization"
method to scattering on graphs was proposed recently in \cite{KS2}.
On the other hand, the mentioned methods are typically used to
evaluate the S--matrix numerically and give little insight, say, into
its dependence on the number $\,N\,$ of the scatterers. To this purpose
a transparent expression for the S--matrix is needed.

Such closed--form formulas were derived recently in
one--dimensional potential scattering, first for an array of
$\,\delta\,$ interactions \cite{BG,Bl,GT} and then for an arbitrary
finitely periodic potential \cite{RRT,SWM}; also the number of
bound states has been discussed in this setting \cite{Ak,SD}. When
this work was in its final stage, another analysis of this
situation appeared \cite{BG} which investigated in detail the
distribution of scattering resonances and the corresponding time
delay. Our basic observation is that the input for the S--matrix
expression are the individual--element scattering data, and thus
the result can be applied to scattering on an array of arbitrary
``black boxes". In the next section we rederive the result of
\cite{RRT,SWM} in this more general context.

We apply the whole scheme to two models of quantum particles confined
to a graph in Sec. 3. The first model consists of an array of
planar loops joined by a pair of external leads. The system is placed
into a homogeneous magnetic field perpendicular to the graph plane.
From several ways how to couple quantum motion on a branching graph
in a self--adjoint way we investigate only $\delta$ coupling.
The next model discussed in this section is made of identical appendices
attached to a line, called comb--shaped graph. The motion on the line is
free, however the particle can interact with an external (scalar) potential
on appendices. We study the most general self--adjoint way of coupling
with the wave function continuous on the line, the only limitation comes
from time--reversal invariance. It leads to a three--parametric family of
solutions. Dependence of transmission probability and resolvent poles on
the number of appendices and the coupling parameters is analyzed and band
spectrum of an infinite comb is discussed as well.

A step more complicated situation is investigated in section 4; scatterers
with dimension two are coupled to single--mode leads. First, we impose
suitable boundary conditions at the junction ensuring self--adjointness and
show their relevancy to the experimentally observed data. Then we specify
our formalism to spheres and find explicite expressions for the transmission
coefficient. We prove existence of resonance peaks and the background which
dominates at large energies and decays not slower than $E^{-1}$ as
$\,E\to\infty\,$. Furthermore, we conjecture that the ``coarse grained"
transmission probability (averaged locally over the resonances) has the
$E^{-1}$ decay typical for the $\,\delta'$ interaction. Finally, we present
band spectrum of infinite system.


\setcounter{equation}{0}
\section{Serial structure transport}

\subsection{Preliminaries}

Consider an equidistant array $\,\{\SS_j:\; j=0,\dots,N\!-\!1\,\}\,$
of identical scatterers placed at the line points
$\,x_0+j\ell\,$. The spacing $\,\ell\ge 0\,$ in this
convention includes only the distance between $\,\SS_j\,$ and
$\,\SS_{j+1}$, not the possible size of the scatterers themselves.

Let us review briefly basic notions concerning a single--mode
scattering on a sole scatterer $\,\SS\,$ placed conventionally at the
point $\,x_0=0\,$. On the two halflines attached to $\,\SS\,$ the
particle moves as free, so the on--shell S--matrix at energy $\,k^2$,
\begin{equation}
\left(\begin{array}{c} B_+ \\ B_- \end{array} \right) \,=\,
S\, \left(\begin{array}{c} A_+ \\ A_- \end{array} \right)\,,
\end{equation}
couples the coefficients of the asymptotic solutions
\begin{equation}
\psi(x)\,=\, \left\lbrace\, \begin{array}{ccc} A_+ e^{ikx}+B_-
e^{-ikx} & \quad\dots\quad & x<0 \\ B_+ e^{ikx}+A_- e^{-ikx} &
\quad\dots\quad & x>0 \end{array} \right.
\end{equation}
In particular, we have
\begin{equation} \label{S}
S\,=\, \left( \begin{array}{cc} t & \tilde r \\ r & \tilde t
\end{array} \right)\,,
\end{equation}
where $\,r,t\,$ and $\,\tilde r,\tilde t\,$ are the left--to--right
and right--to--left reflection and transmission amplitudes,
respectively. We shall consider only the non--dissipative situation
when $\,S\,$ is unitary; its dependence on the momentum $\,k\,$
will be indicated only if necessary.

Since the interaction responsible for the scattering is localized by
assumption, solutions to the Schr\"odinger equation acquire the
asymptotic form outside the scatterer. Hence $\,S\,$ may be expressed
alternatively in terms matrices used in the theory of ordinary
differential equations. One is the ``coefficient" transfer matrix
$\,M\,$ relating the solutions to the left and to the right of the
scatterer by
\begin{equation}
\left(\begin{array}{c} B_+ \\ A_- \end{array}\right)
\,=\, M\, \left(\begin{array}{c} A_+ \\ B_- \end{array}\right)\,.
\end{equation}
It is straightforward to see that
\begin{equation} \label{M}
M\,=\,{1\over\tilde t}\, \left(\begin{array}{cc}
t\tilde t \!-\! r\tilde r & \:\tilde r \\
-r & \:1 \end{array}\right)\,,
\end{equation}
and vice versa,
\begin{equation} \label{M-rt}
r=\,-\,{M_{21}\over M_{22}}\,,\qquad t = M_{11} -\,
{M_{12}M_{21}\over M_{22}}\,, \qquad \tilde{r} =\,
{M_{12}\over M_{22}}\,,\qquad \tilde t=\,{1\over M_{22}}\,.
\end{equation}
Another one is the transfer matrix which relates the values and
derivatives of the two solutions,
\begin{equation} \label{L}
\left(\begin{array}{c} u(0+) \\ u'(0+) \end{array}\right) \,=\, L\,
\left(\begin{array}{c} u(0-) \\ u'(0-) \end{array}\right)\,.
\end{equation}
Substituting the boundary values of $\,u(x)= e^{ikx}+r e^{-ikx}$ for
$\,x<0\,$ and $\,u(x)= te^{ikx}$ for $\,x>0\,$ into this relation, we
get a pair of equations for $\,r,\,t\,$ which is solved by
\begin{eqnarray} \label{rt}
r &\!=\!&  -\,\frac{L_{21} + ik(L_{22} \!-\! L_{11}) + k^2L_{12}}
{L_{21} - ik(L_{22} \!+\! L_{11}) - k^2L_{12}}\,, \nonumber \\ \\
t &\!=\!& -\,\frac{2ik\,\det{L}}{L_{21} - ik(L_{22} \!+\! L_{11}) -
k^2L_{12}} \nonumber
\end{eqnarray}
In the same way we get the right--to--left amplitudes,
\begin{eqnarray} \label{tilde rt}
\tilde r &\!=\!&  -\,\frac{L_{21} + ik(L_{11} \!-\! L_{22}) +
k^2L_{12}} {L_{21} - ik(L_{22} \!+\! L_{11}) - k^2L_{12}}\,,
\nonumber \\ \\
\tilde t &\!=\!& -\,\frac{2ik}{L_{21} - ik(L_{22} \!+\! L_{11}) -
k^2L_{12}} \nonumber
\end{eqnarray}
Combining (\ref{M}) with (\ref{rt}), (\ref{tilde rt}) we can express
the ``coefficient" transfer matrix as
\begin{equation} \label{M-L}
M\,=\, \frac{1}{2ik}\left(\begin{array}{cc}
L_{21}+ik(L_{11}\!+\!L_{22})-k^2L_{12} &
L_{21}+ik(L_{11}\!-\!L_{22})+k^2L_{12} \\ \\
-L_{21}+ik(L_{11}\!-\!L_{22})-k^2L_{12} &
-L_{21}+ik(L_{11}\!+\!L_{22})+k^2L_{12}
\end{array}\right)\,.
\end{equation}
The class of admissible transfer matrices is restricted by the
S--matrix unitarity. In particular, the conservation of probability
current, $\,|r|^2\!+|t|^2=|\tilde r|^2\!+|\tilde t|^2=1\,$ implies
$\,|\det{L}|=1\,$; we shall write therefore
\begin{equation} \label{det L}
\det{L}\,=\; {t\over\tilde t}\;=:\, e^{2i\varphi}\,.
\end{equation}
One can also express $\,L\,$ by means of the other two matrices.
For instance, suppose that (\ref{S}) is given. Using the relation
(\ref{L}) with the boundary values for the left--to--right scattering,
the same for the right--to--left case with the explicitly written
$\,L^{-1}$ and $\, t= \tilde t\det L\,$, we get a system of four linear
equations for $\,L_{jk}$,
\begin{eqnarray} \label{L system}
L_{11}(1\!+\!r)+ikL_{12}(1\!-\!r)\,=\,t\,, &&
L_{21}(1\!+\!r)+ikL_{22}(1\!-\!r)\,=\,ikt\,, \nonumber \\ \\
L_{22}(1\!+\!\tilde r)+ikL_{12}(1\!-\!\tilde r)\,=\,t\,, &&
L_{21}(1\!+\!\tilde r)+ikL_{11}(1\!-\!\tilde r)\,=\,ikt\,. \nonumber
\end{eqnarray}
Only three of them are independent; it is straightforward to see
that the first with the third, and the second with the fourth
equation lead to the same relation which is solved by
$$
L_{22}\,=\, L_{11}\, {(1\!+\!r)(1\!-\!\tilde r)\over
(1\!-\!r)(1\!+\!\tilde r)}\,-\, {t(r\!-\!\tilde r) \over
(1\!-\!r)(1\!+\!\tilde r)}\,.
$$
The same pairs of equations allow us to express $\,L_{12}$ and
$\,L_{21}$ in terms of $\,L_{11}$ and $\,L_{22}$; in combination
with the last relation we find
$$
L_{12}\,=\, -\,L_{11}\, {(1\!+\!r)\over ik(1\!-\!r)}\,+\,
{t\over ik(1\!+\!\tilde r)}\,, \qquad
L_{21}\,=\, -ikL_{11}\, {(1\!-\!\tilde r)\over ik(1\!+\!\tilde r)}
\,+\, {ikt\over (1\!+\!\tilde r)}\,.
$$
We have still the condition (\ref{det L}). Computing the determinant
with the help of the above relations, we get
$$
\det L\,=\, {t(2L_{11}\!-t)\over (1\!-\!r)(1\!+\!\tilde r)} \,=\,
{t\over\tilde t}\,.
$$
One can express $\,L_{11}\,$ from here and substitute into the
formulas for the other elements; this yields finally
\begin{equation} \label{L-S}
L\,=\, {1\over 2\tilde t}\,
\left( \begin{array}{cc} t\tilde t+(1\!+\!r)(1\!-\!\tilde r) &\quad
{1\over ik}\, \left\lbrack t\tilde t-(1\!+\!r)(1\!+\!\tilde r)
\right\rbrack \\ \\
ik \left\lbrack t\tilde t-(1\!-\!r)(1\!-\!\tilde r) \right\rbrack
&\quad t\tilde t+(1\!-\!r)(1\!+\!\tilde r) \end{array} \right)\,.
\end{equation}
The determinant of this matrix is equal to the middle expression of
(\ref{det L}) and substituting into (\ref{rt}), (\ref{tilde rt}) one
can check that (\ref{L-S}) indeed represents the inverse transformation.

Furthermore, the unitarity of $\,S\,$ has a stronger consequence. It
is well known that a general $\,2\times 2\,$ unitary matrix can be
parametrized by four real numbers as
$$
e^{i\xi} \left( \begin{array}{cc} e^{i(\alpha+\delta)}\cos\beta &
e^{i(\delta-\alpha)}\sin\beta \\ -e^{i(\alpha-\delta)}\sin\beta &
e^{-i(\alpha+\delta)}\cos\beta \end{array}\right)\,.
$$
Using this for $\,S\,$ given by (\ref{S}) and substituting into
(\ref{L-S}) we find that
\begin{equation} \label{L phase}
L\,=\, e^{i\varphi} \LL\,, \qquad \LL\;\;{\rm real\;with\;\quad}
\det\LL=1 \,,
\end{equation}
where $\,\varphi:=\alpha\!+\!\delta\,$. Notice that $\,M\,$ given by
(\ref{M-L}) has then the following property:
\begin{equation} \label{M phase}
\overline{M}_{11}\,=\, e^{-2i\varphi} M_{22}\,, \qquad
\overline{M}_{12}\,=\, e^{-2i\varphi} M_{21}\, .
\end{equation}
%


\subsection{Recursive relations for scattering amplitudes}

Before we derive the mentioned closed--form expression, let us recall
the usual factorization technique. We index the transmission and
reflection amplitudes for the array by $\,N\,$. In analogy with
(\ref{S}) we have
\begin{equation}\label{S_N}
\left(\begin{array}{c} B_-^N \\ B_+^N
\end{array}\right) \,=\,
\left(\begin{array}{cc} r_N & \tilde t_N \\
t_N & \overline\eps^{2(N-1)}\tilde r_N
\end{array}\right)
\left(\begin{array}{c} A_+^N \\ A_-^N \end{array}\right)\,,
\end{equation}
where $\,\eps:= e^{ik\ell}$. Next we add the $\,(N\!+\!1)$--th
scatterer to the right side of the array for which
\begin{equation}\label{S_add}
\left(\begin{array}{c}
B_- \\ B_+ \end{array}\right) \,=\,
\left(\begin{array}{cc} \eps^{2N}r & \tilde t \\
t & \overline\eps^{2N}\tilde r \end{array}\right)
\left(\begin{array}{c} A_+ \\ A_- \end{array}\right)\,,
\end{equation}
where, of course, $\,B_+^N = A_+\,$ and $\,B_- = A_-^N$. In analogy
with (\ref{M}) we rewrite the last two relations in the
``coefficient" transfer matrix form. Multiplying the two matrices we
get $\,{B_+ \choose A_-}=M{A_+^N \choose B_-^N}\,$ with
$$
M\,:=\, \left( \begin{array}{cc} \frac{1}{\tilde t\tilde
t_N}((t\tilde t\!-\!r\tilde r)( t_N\tilde t_N
\!-\!r_N\tilde r_N \bar\eps^{2(N-1)}) \!-\! r_N\tilde r\bar \eps^{2N}) &
\frac{1}{\tilde t\tilde t_N}((t\tilde t\!-\!r\tilde r)\tilde r_N
+\bar \eps^{2N}\tilde r)  \\ \\
\frac{1}{\tilde t \tilde t_N}(-\eps^{2N}r(t_N\tilde t_N\!-\!r_N
\tilde r_N)\!-\! r_N) & \frac{1}{\tilde t\tilde t_N}(1
-\eps^{2N}r\tilde r_N)
\end{array} \right)\,.
$$
Comparing this with (\ref{M-rt}) we find the sought recursive
relations
\begin{equation} \label{tilde recurs}
\tilde r_{N+1} \,=\,
\overline\eps^{2N}\tilde r + \,\frac{\tilde r_N\tilde tt}{1 -
\eps^{2N}\tilde r_Nr}\,,\qquad
\tilde t_{N+1} \,=\, \frac{\tilde t\tilde t_N}{1 -
\eps^{2N}r\tilde r_N}\,.
\end{equation}
If we modify the argument by adding the $\,(N\!+\!1)$--th scatterer
to the left of the array, we get in the same way
\begin{equation} \label{recurs}
r_{N+1} \,=\,
\eps^{2N}r + \frac{r_N\tilde tt}{1 -
\overline\eps^{2N}\tilde rr_N}\,,\qquad
t_{N+1} \,=\, \frac{tt_N}{1 - \overline\eps^{2N}r_N\tilde r}\,.
\end{equation}
Since the ``component" S--matrices are unitary, it is straightforward
to check by induction that the same is true for the total S--matrix.
The relations (\ref{tilde recurs}) and (\ref{recurs}) have a
transparent meaning: expanding the fractions into geometric series we
obtain expressions containing sums of contributions from various
scattering processes.


\subsection{The S--matrix expression}

On the other hand, the recursive expressions do not relate directly
the S--matrices of an individual scatterer and that of the whole array.
To this end, notice first that in view of (\ref{M phase}) we can write
the $\,M\,$ matrix of the $\,j$--th scatterer as
\begin{equation} \label{M_j}
M_j\,=\, e^{i\varphi}\, \left(\begin{array}{cc}
\overline R & \overline\eps^{2j}\overline S \\
\eps^{2j}S & R \end{array}\right),
\end{equation}
where $\,\eps := e^{ik\ell}$ as above and
\begin{eqnarray} \label{RS}
R &\!:=\!& e^{-i\varphi} M_{22}\,=\, {e^{-i\varphi}  \over\tilde t}
\,=\,{e^{i\varphi} \over t} \,=\,\frac{\LL_{11}+\LL_{22}}{2}\,
+i\left(\frac{\LL_{21}}{2k}-\frac{k}{2}\LL_{12} \right)\,, \nonumber
\\ \\
S &\!:=\!& e^{-i\varphi} M_{21}\,=\, -\,{e^{-i\varphi}r \over\tilde t}
\,=:\,-\,{e^{i\varphi}r \over t}\,=\, \frac{\LL_{11}-\LL_{22}}{2}\,
+i\left(\frac{\LL_{21}}{2k}+\frac{k}{2}\LL_{12} \right)\,. \nonumber
\end{eqnarray}
By definition the ``coefficient" transfer matrix of the array is
obtained by multiplying successively the matrices (\ref{M_j}). We
denote $\,M^{(n)} := M_nM_{n-1}\dots M_{0}\,$. It is easy to compute
the first few matrices $\,M^{(n)}$; this inspires us to look for the
general product in the form
\begin{equation}
M^{(n)} \,=\, e^{i(n+1)\varphi}\,
\left(\begin{array}{cc}
\overline\eps^{n+1}\vert S\vert^{n+1}\overline\gamma_n &
\overline\eps^n\overline S\vert S\vert^n\overline\delta_n \\
\eps^n S\vert S\vert^n\delta_n &
\eps^{n+1}\vert S\vert^{n+1}\gamma_n
\end{array}\right)\,,
\end{equation}
where the coefficients have to satisfy the recursive relations
\begin{equation}\label{coef recurs}
\gamma_{n+1} \,=\, \zeta\gamma_n + \overline\delta_n\,, \qquad
\delta_{n+1} \,=\, \zeta\delta_n + \overline\gamma_n
\end{equation}
with $\,\delta_0 = 1\,$ and
\begin{equation}\label{z}
\gamma_0\,=\, \zeta\,:=\, \frac{\overline\eps R}{\vert S\vert}\,,
\end{equation}
which follows from $\,M^{(n+1)} = M_{n+1}M^{(n)}\,$. Since
\begin{equation} \label{det M_j}
\det M_j \,=\,e^{2i\varphi} \left(|R|^2- |S|^2\right) \,=\,
e^{2i\varphi}\, {1-|r|^2\over |\tilde t|^2}\,=\, e^{2i\varphi}\,,
\end{equation}
and consequently, $\,\det M^{(n)}\,=\, e^{2i(n+1)\varphi} |S|^{2n+2}
(|\gamma_n|^2\!- |\delta_n|^2) \,=\, e^{2i(n+1)\varphi}$, we have
\begin{equation} \label{gamma}
\gamma_n \,=\, e^{i\theta_n}\,\sqrt{|\delta_n|^2+| S|^{-2n-2}}
\end{equation}
with a phase factor to be determined. Substituting into the relations
(\ref{coef recurs}) we get
\begin{eqnarray}\label{coef recurs 2}
e^{i\theta_{n+1}}\, \sqrt{|\delta_{n+1}|^2+|S|^{-2n-4}} &\!=\!&
\zeta\, e^{i\theta_n}\, \sqrt{|\delta_n|^2+|S|^{-2n-2}}\,
+\delta_n\,, \nonumber \\ \\
\delta_{n+1} &\!=\!& \zeta\delta_n+\,e^{-i\theta_n}\,
\sqrt{|\delta_n|^2+|S|^{-2n-2}}\,. \nonumber
\end{eqnarray}
We express $\,e^{i\theta_n}$ from the second equation and substitute
into the first one; this yields
\begin{equation}\label{coef recurs 3}
\delta_{n+2}-(\zeta+\overline\zeta)\delta_{n+1}+(|\zeta|^2-1)
\delta_n \,=\,0\,.
\end{equation}
Now $\,\delta_0=1\,$ and $\,\delta_1 = \zeta +\overline\zeta\,$,
so (\ref{coef recurs 3}) is solved by
$$
\delta_n\,=\, \left(|\zeta|^2\!-1\right)^{n/2}\, U_n\left(\frac{\zeta
+\overline\zeta}{2\sqrt{|\zeta|^2\!-1}}\right)\,,
$$
where $\,U_n\,$ is the Chebyshev polynomial of the second kind. Since
$\,|\zeta|^2\!-1 = |S|^{-2}\,$ by (\ref{det M_j}), its argument
can be more compactly written as $\,\re(\bar\eps R)\,$.
Using (\ref{gamma}) and (\ref{coef recurs 2}) again, we find
$\,M^{(n)}\,=\, e^{2i(n+1)\varphi} \MM^{(n)}$ with
\begin{equation}
\MM^{(n)}=\, \left(\begin{array}{cc}
\overline\eps^{n+1}e^{-i\theta_n}\sqrt{1+|S|^2 U_n(\re(\overline\eps
R))^2} & \!\!\overline\eps^n\overline SU_n(\re(\overline\eps R)) \\ \\
\eps^n SU_n(\re(\overline\eps R)) & \!\!\eps^{n+1}e^{i\theta_n}
\sqrt{1+|S|^2 U_n(\re(\overline\eps R))^2} \end{array}\right)\,,
\end{equation}
where
\begin{equation} \label{serial phase}
e^{i\theta_n} \,=\, {U_{n+1}(\re(\overline\eps R)) -\eps\overline
RU_n(\re(\overline\eps R))\over \sqrt{1+|S|^2 U_n^2(\re(\overline\eps
R))}}\,.
\end{equation}
Now we may employ (\ref{M}) and (\ref{det L}) to find the sought
formulas for the array of $\,N\,$ scatterers; it is sufficient
to put $\,n=N-1\,$. We arrive at the following conclusion:
\vspace{2mm}

\begin{theorem}
With the given notation, the transmission and reflection amplitudes
of an $\,N$--element serial structure express as
\begin{eqnarray}
t_N &\!=\!& {\overline\eps^N\, e^{-i\theta_{N-1}} \over
\sqrt{1+|S|^2 U_{N-1}(\re(\overline\eps R))^2}}  \label{t_N} \\
\nonumber \\
r_N &\!=\!& -\, {\overline\eps\,
e^{-i\theta_{N-1}}\,SU_{N-1}(\re(\overline\eps R)) \over
\sqrt{1+|S|^2 U_{N-1}(\re(\overline\eps R))^2}}\,, \label{r_N}
\end{eqnarray}
where the phase factor is given by (\ref{serial phase}). In the same
way the right--to--left amplitudes are $\,\tilde t_N= t_N
e^{-2iN\varphi}$ and
$$
\tilde r_N \,=\, -\, {\overline\eps^{2N-1} e^{-i\theta_{N-1}}\,
\overline S U_{N-1}(\re(\overline\eps R)) \over \sqrt{1+|S|^2
U_{N-1}(\re(\overline\eps R))^2}}\,.
$$
In particular, the transmission and reflection probabilities are the
same in both directions and equal
\begin{equation} \label{t_N^2}
|t_N|^2=\, {1 \over 1+|S|^2U_{N-1}(\re(\overline\eps R))^2}\,,
\quad |r_N|^2=\, {|S|^2 U_{N-1}(\re(\overline\eps R))^2 \over
1+|S|^2 U_{N-1}(\re(\overline\eps R))^2}\,.
\end{equation}
\end{theorem}
\vspace{3mm}

\noindent
Recall that
$$
|S|^2\,=\, \left| r\over t\right|^2\,, \qquad \re(\overline\eps
R)\,=\, \re\left(e^{-i(k\ell+\varphi)} \over\tilde t \right)\,=\,
\re\left(e^{-i(k\ell-\varphi)} \over t \right)\;;
$$
it is obvious from (\ref{t_N^2}) that the probability current is
preserved.


\subsection{Relation to band spectra of periodic systems}

Consider now an infinite periodic array of identical scatterers
$\,\SS\,$ joined by line segments of length $\,\ell\,$. The
one--period transfer matrix is $\,T=L\,{\cal U}_\ell(k)\,$, where
$$
{\cal U}_\ell(k)\,:=\, \left( \begin{array}{cc} \cos k\ell &
{1\over k} \sin k\ell \\ -k\sin k\ell & \cos k\ell
\end{array} \right)
$$
coresponds to the segment. The band spectrum of the problem is given
by the Bloch condition, $\,\det\left(T-e^{i\theta}\right)=0\,$, or
\begin{equation}
e^{2i\theta} -e^{i\theta}\tr T+ \det T\,=\,0\,.
\end{equation}
In view of (\ref{det L}), $\;\det T= \det L= e^{2i\varphi}$, so the
condition may be written as
$$
e^{-i\varphi} \tr T\,=\, 2\cos(\theta\!-\!\varphi)\,.
$$
To express the \lhs, we employ (\ref{L-S}) which yields
$$
\tr T\,=\, \tr T\, \cos k\ell\,+\, \left({1\over k}L_{21}- kL_{12}
\right)\,\sin k\ell\,=\,
{t\tilde t-r\tilde r\over \tilde t}\, e^{ik\ell}\, +\,
{1\over \tilde t}\, e^{-ik\ell}\,,
$$
and since
$$
{t\tilde t-r\tilde r\over \tilde t}\,=\,{e^{2i\varphi}\over
\overline{\tilde t}}
$$
by (\ref{M}) and (\ref{M phase}), we arrive finally at
\begin{equation} \label{Bloch}
\re(\overline\eps R)\,=\, {e^{i(k\ell+\varphi)} \over
2\overline{\tilde t}}\,+\, {e^{-i(k\ell+\varphi)} \over 2\tilde t}
\,=\, \cos(\theta\!-\!\varphi)\,.
\end{equation}
The \lhs as a function of $\,k\,$ is typically oscillating. Since the
amplitude $\,|t|^{-1}>1\,$ unless the single--element scattering is
reflectionless, the periodic spectrum has gaps in general.

The relations (\ref{t_N^2}) show how the band spectrum arises in the
limit $\,N\to\infty\,$ of the serial--structure scattering. The
Chebyshev polynomials
\begin{equation}
U_n(x)\,=\, \sum_{m=0}^{[n/2]} (-1)^m\, {(n\!-\!m)!\over m!(n\!-\!2m)!}\,
(2x)^{n-2m}
\end{equation}
are oscillating within the interval $\,[-1,1]\,$. The easiest way to
see that is to use the representation
\begin{equation} \label{band Chebyshev}
U_n(x)\,=\, {\sin (n\!+\!1)\xi \over \sin\xi}\,, \qquad \xi:=
\arccos x\,.
\end{equation}
Thus $\,U_{N-1}\,$ has $\,N-1\,$ roots in $\,[-1,1]\,$ and each band
contains at least $N-1$ points where $\,|t_N(k)|^2=1\,$. Possible
additional points with this property can come from zeros of the
single--element reflection coefficient. Properties of Chebyshev
polynomials yield also lower bounds to the envelope of the
transmission probability oscillations. The representation (\ref{band
Chebyshev}) implies $\,|U_n(x)|\le (1\!-\!x^2)^{-1/2}\,$ for
$\,|x|<1\,$, and therefore
\begin{equation} \label{lb1 inside}
|t_N(k)|^2\,\ge\, {1-(\re(\overline\eps R))^2 \over 1+|S|^2
-(\re(\overline\eps R))^2} \,=\, {|t(k)|^2-(\re(\tilde t(k)
e^{i(k\ell+ \varphi)}))^2 \over 1-(\re(\tilde t(k)
e^{i(k\ell+ \varphi)}))^2}\,.
\end{equation}
If $\,|t(\cdot)|^2\,$ is a slowly varying function, the \rhs reaches
its maximum $\,|t(k)|^2\,$ in the middle of the band; it is zero at
the band edges. However, the transmission can vanish within a band
only due to a single--element full reflection. This follows from
another upper bound \cite{ASt}, $\;|U_n(x)|\le (n\!+\!1)\,$ for
$\,|x|\le 1\,$, which yields $\,|t_N|^2\ge (1+N^2|S|^2)^{-1}$ or
\begin{equation} \label{lb2 inside}
|t_N(k)|^2\,\ge\, {|t(k)|^2\over 1+(N\!-\!1)|r(k)|^2}\,.
\end{equation}

On the other hand, to see the behavior of the reflection and
transmission amplitudes in the gaps of the periodic spectrum, we have
to estimate the Chebyshev polynomials outside $\,[-1,1]\,$. By
analytical continuation, the relation (\ref{band Chebyshev}) gives
$\,U_n(x)= \sinh \left((n\!+\!1){\rm arcosh\,} x\right) / \sinh
\left( {\rm arcosh\,} x\right) \,$, so
$$
U_n(x)\,=\, \sum_{k=0}^{n} \left( x+\sqrt{x^2\!-1} \right)^{n-2k}\,.
$$
Then we have, for instance, the following estimates
\begin{equation} \label{gap estimates}
n+x^n \,\le\, U_n(x)\,\le\, (n\!+\!1)\,\left( x+\sqrt{x^2\!-1}
\right)^n\,.
\end{equation}
The first inequality yields an upper bound,
\begin{equation} \label{ub outside}
|t_N(k)|^2 \,\le\, {1\over 1+ \left( N\!-\!1\!+ (\re(\overline\eps
R))^{N-1} \right)^2}\,.
\end{equation}
It is clear that $\,|t_N(k)|^2=1\,$ holds only if the same is true
for $\,|t(k)|^2$; in all the other cases it behaves as $\,o\left(
\left(\re(\overline\eps R) \right)^{2N-2}\right)\,$ as
$\,N\to\infty\,$. On the other hand, $\,|t_N(k)|^2=0\,$ holds only if
a single scatterer has a full reflection at this energy, otherwise
the second inequality of (\ref{gap estimates}) together with
$\,|\re(\overline\eps R)| \le |t(k)|^{-1}$ and the unitarity relation
give
\begin{equation} \label{lb outside}
|t_N(k)|^2 \,\ge\, {|t(k)|^{2N} \over |t(k)|^{2N}+ N^2 |r(k)|^2
(1\!+\!|r(k)|)^{2N-2}} \,.
\end{equation}


\setcounter{equation}{0}
\section{Serial graphs}

As we have said we want now to illustrate the above results on
several examples which go beyond the usual one--dimensional potential
scattering. In this section we shall discuss the situation the serial
structure is a graph.

Although nonrelativistic quantum mechanics for quantum particles
confined to a graph has been considered already several decades ago
in connection with the free--electron models of hydrocarbons
\cite{RuS}, it became a subject of intense interest only recently as
a tool to describe systems of quantum wires --- see
\cite{Ad,AL,AEL,ARZ,Bu,E1,E2,E3,E4,EG,ES1,ESe,GP,GLRT,JS,KS1} and
references therein. There are also other systems for which graph
description could prove to be useful such as objects composed of carbon
nanotubes \cite{CCB,HSO}.

Graph systems are attractive because they are often explicitly
solvable; on the mathematical level they represent systems of
ordinary differential equations in contrast to partial differential
equations with nontrivial boundary conditions needed to describe
quantum wire systems in higher dimensions. At the same time, the
simplified description may still preserve basic features of the real
system. Of course, replacing a branched waveguide system by its
skeleton graph is a nontrivial approximation, but we shall avoid
discussing this point here; some comments can be found in \cite{E2,KZ}
and references given there.


\subsection{Loop arrays}

In our first example an individual scatterer is a planar loop
$\,\LL\,$ with a pair of external leads placed into a homogeneous
magnetic field --- \cf Fig.~1.
\begin{figure}
\setlength\unitlength{1mm}
\begin{picture}(125,45)(10,15)
\thicklines
\put(48,30){\line(1,0){30.5}}
\put(92.5,30){\line(1,0){30}}
\put(85.5,30){\circle{25}}
\thinlines
\put(85.5,30){\circle{3}}
\put(85.5,30){\circle*{.8}}
\put(87,28){\mbox{$B$}}
\put(85.5,23){\vector(1,0){10}}
\put(96,21){\mbox{$A$}}
\put(58,35){\mbox{$f_1$}}
\put(58,25){\mbox{${\R}^-$}}
\put(110,35){\mbox{$f_2$}}
\put(110,25){\mbox{${\R}^+$}}
\put(82,40){\mbox{$[0,L_2]$}}
\put(82,18){\mbox{$[0,L_1]$}}
\put(84,25){\mbox{$u_1$}}
\put(84,34){\mbox{$u_2$}}
\end{picture}
\caption{A loop--graph scatterer in a magnetic field}
\end{figure}
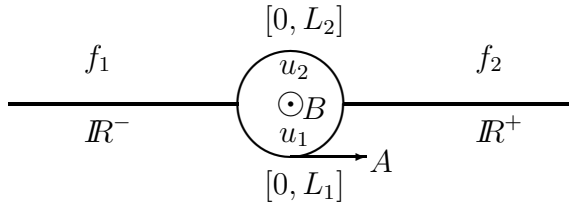
We suppose that the field is perpendicular to the graph plane. The
corresponding vector potential in the circular gauge is $\,\vec
A(\vec x)= \left( -{1\over 2}By, {1\over 2}Bx,0 \right)\,$ and
$$
\int_{\LL} \vec A(\vec x)\, d\vec x\,=\, {BS\over L_1\!+L_2} \,=\,
{\Phi\over L_1\!+L_2}\,,
$$
where $\,S\,$ is the loop area, $\,L_j\,$ are the lengths of its two
branches, and $\,\Phi\,$ is the corresponding magnetic flux.

Such a scattering system has been considered earlier --- see
\cite{AS,Bu,JS} and references therein. The Hilbert space for the
graph of Fig.~1 is the orthogonal sum of four $\,L^2$ spaces
referring to the graph links; its elements will be denoted as
$\,(f_1,u_1,u_2,f_2)\,$ with the coordinates at the loop taken
anticlockwise. We suppose that the particle has a unit charge and
apart of the magnetic field it moves as free on the graph. There
are various ways how to couple the operators $\,\left(
-i\partial_x -A\right)^2$ from different graph links in a
self--adjoint way \cite{ES2,ESe,KS1}. For the sake of simplicity we
restrict ourselves to the usual $\,\delta\,$ coupling, \ie, we
impose the boundary conditions
\begin{eqnarray} \label{loop bc}
f_1(0)=u_1(0)=u_2(L_2)\, && -f'_1(0)+u'_1(0)-u'_2(L_2)=
\alpha_1f_1(0)\,, \nonumber \\
f_2(0)=u_2(0)=u_1(L_1)\, && \phantom{A} f'_2(0)-u'_1(L_1)+u'_2(0)=
\alpha_2f_2(0)\,,
\end{eqnarray}
where the function and derivative values mean the appropriate
one--sided limits. On the other hand, we do not suppose in general
that $\,\alpha_1,\,\alpha_2\,$ and $\,L_1,\,L_2\,$ are the same.

Substituting into (\ref{loop bc}) the boundary values of the
generalized eigenfunctions on the loop, $\,u_j(x)= A_j e^{-i(A-k)x}\!
+ B_j e^{-i(A+k)x},\; j=1,2\,$, we get a system of six equations.
Eliminating from here $\,A_j,\,B_j\,$ we arrive at the relations
\begin{eqnarray*}
f_2(0) &\!=\!& \frac{1}{2kD}\left((i\alpha_1B\!+\!2kC)f_1(0)
+iBf'_1(0) \right) \\ \\
f'_2(0) &\!=\!& \frac{1}{2kD}\left((i\alpha_1\alpha_2B
+2(\alpha_1\!+\!\alpha_2)kC-4ik^2E_-E_+)f_1(0) +(2kC\!+\!i\alpha_2B)
f'_1(0)\right)\,,
\end{eqnarray*}
determining the transfer matrix, where
\begin{eqnarray*}
B &\!=\!& (\eps_{1+}\!-\!\eps_{1-})(\eps_{2-}\!-\!\eps_{2+})
\,=\,4e^{-i\Phi}\sin{kL_1}\sin{kL_2}\,, \\
C &\!=\!& -\eps_{1-}\eps_{2-}+\eps_{1+}\eps_{2+}
\,=\, 2ie^{-i\Phi}\sin{k(L_1\!+\!L_2)}\,, \\
D &\!=\!& \eps_{2+}-\eps_{2-}+\eps_{2+}\eps_{2-}
(\eps_{1+}\!-\!\eps_{1-})\,, \\
E_{\pm} &\!=\!& 1-\eps_{1\pm}\eps_{2\pm} \,=\, 2i\, e^{-i\Phi/2}
e^{\pm ik(L_1+L_2)/2} \sin\left( \pm k(L_1+L_2)- \Phi \over 2
\right)\,, \\
\eps_{j\pm} &\!=\!& e^{i(-A\pm k)L_j}\,,\quad j=1,2.
\end{eqnarray*}
The reflection and transmission amplitudes are given by
\begin{eqnarray}
r(k) &\!=\!&
\frac{-i\alpha_1\alpha_2B+k((\alpha_2\!-\!\alpha_1)B-
2(\alpha_1\!+\!\alpha_2)C) +ik^2(4E_-E_+\!-B)}
{i\alpha_1\alpha_2B+k(\alpha_2\!+\!\alpha_1)(2C+B)
-ik^2(4E_-E_+\!+B+4C)}\,, \nonumber \\ \\
t(k) &\!=\!& \frac{4ik^2D e^{-2i\Phi}} {i\alpha_1\alpha_2B
+k(\alpha_2\!+\!\alpha_1)(2C+B)-ik^2(4E_-E_+\!+B+4C)}\,, \nonumber
\end{eqnarray}
respectively. Since $\,B,\,C,\,E_{\pm}$ as well as $D^2$
%
%
are $\,2\pi$--periodic functions of the magnetic flux, the same is
valid for the reflection and transmission {\em probabilities.} Recall that
if we put $e = \hbar = c = 1$, then $\,2\pi$ is the magnetic flux quantum
in these units.


\subsection{Band spectrum of an infinite loop array}

We illustrate the relation of transmission probabilities of a finite array
of loops and the spectrum of the corresponding infinite system on Fig.~2.
As already mentioned $|t_N(k)|^2$ is a $\,2\pi$--periodic function and
because the condition (\ref{Bloch}) determining the band spectrum of the
infinite system can be reformulated as $0 < |\re(\overline\eps R)| < 1$
we show dependence on $\Phi$ only in the range $[0,2\pi]$
($|\re(\overline\eps R)|$ is also a $\,2\pi$--periodic function). We choose
loops with different $L_1, L_2$ and $\alpha_1, \alpha_2$. Even for a relatively
small number of loops $N = 6$ the values of $|t_N(k)|^2$ are clearly nonzero
in areas of parameters $\Phi$ and $k$ where there are bands of the infinite
system and negligible where there are gaps.


\subsection{Comb graphs}

Our next example concerns the case of a comb--shaped graph, \ie,
a line with a finite number $\,N\,$ of identical appendices attached
to it at equally spaced points. Such systems have been discussed
recently \cite{SJ} following earlier studies of a single--stub
waveguide \cite{PZL,SMRH1,SMRH2,TB}.

Comparing to the previous work and the preceding example, we shall
discuss comb--shaped graphs in a more thorough way. First of all,
instead of the $\,\delta$--coupling used above (or the Griffith's
boundary conditions in the terminology of \cite{SJ}) we allow for
the most general self--adjoint way in which the stubs can be attached
to continuous wavefunctions on the line. This amounts to imposing at
the junctions the boundary conditions adopted from an earlier treatment
of the T--shaped graph \cite{ESe}. Should the Hamiltonian be
time--reversal invariant, the junction is then characterized by three
real parameters. In this framework we are able to handle imperfect
contacts \cite{E3}; moreover, it is straightforward to modify the
results derived  below to graphs with a $\,\delta'$--coupling which
corresponds to the situation where the junction itself represents a
complicated geometric scatterer (see \cite{AEL,E2}, more about that will
be said in the next section). Computing the S--matrix we also assume
that the particle is under influence of a potential on the stubs; this
makes it possible to investigate how the band--form zones of high
transmission which arise for $\,N\gg 1\,$ change when an external field
is applied.

The formula derived in the previous section allows us to express the
transmission and reflection probabilities. In addition to them, one is
able to generalize the result of \cite{ESe} to the present situation
and to find an explicit Krein--formula expression for the resolvent of
the comb--graph Hamiltonian. This allows us to study their resonance
structure of the problem which arises from perturbation of the
disconnected--stub discrete spectrum embedded into the continuum of
the line motion. The mentioned expression yields an equation from
which the resolvent singularities on the second sheet can be found.

After this introduction, let us describe the model. For a greater
generality we suppose first that the appendices are not
necessarily identical. The graph $\,\Gamma_N\,$ will therefore
consist of a line with a finite sequence
$\,\{(s\!-\!1)\ell\}_{s=1}^N\,$ of points at which appendices of
finite lengths $\,L_s\,$ are attached (see Fig.3).
\setcounter{figure}{2}
\begin{figure}
\setlength\unitlength{1mm}
\begin{picture}(150,40)
\linethickness{.6mm}
\put(15,10){\line(1,0){60}}
\put(80,10){\dots}
\put(90,10){\line(1,0){45}}
\put(40,10){\line(0,1){20}}
\put(55,10){\line(0,1){20}}
\put(70,10){\line(0,1){20}}
\put(95,10){\line(0,1){20}}
\put(110,10){\line(0,1){20}}
\thinlines
\put(40,30){\line(-1,0){10}}
\put(32,20){\vector(0,1){10}}
\put(32,20){\vector(0,-1){10}}
\put(33,20){$L_1$}
\put(110,30){\line(-1,0){10}}
\put(102,20){\vector(0,1){10}}
\put(102,20){\vector(0,-1){10}}
\put(103,20){$L_N$}
\put(40,10){\line(0,-1){10}}
\put(55,10){\line(0,-1){10}}
\put(45,2){\vector(-1,0){5}}
\put(45,2){\vector(1,0){10}}
\put(47,3){$\ell$}
\end{picture}
\caption{A comb--shaped graph}
\end{figure}
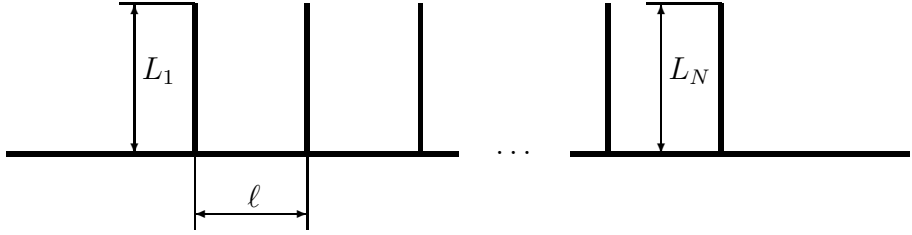
The state Hilbert space of the problem is then $\,\HH\equiv
L^2(\Gamma_N):= L^2(\R)\oplus \left( \bigoplus_{s=1}^N L^2(0,L_s)
\right)\;$; we shall write its elements as columns
$$
\psi \,=\, \left(\begin{array}{c}
f \\ u_1\\ \vdots \\ u_N \\ \end{array}\right)\,.
$$
We suppose that the motion at the backbone line is free while the
particle is exposed to potentials $\,V_s\,$ on the ``teeth"; hence
the Hamiltonian of a nonrelativistic particle of mass $\,m=1/2\,$
living on $\,\Gamma_N\,$ acts as
\begin{equation} \label{comb Hamiltonian}
(H\psi)_1(x) := -f''(x)\,,\qquad (H\psi)_{s+1}(x) :=
(-u_s''\!+V_su_s)(x)\,,\quad s=1,\dots,N\,,
\end{equation}
out of the junctions. To make it a self--adjoint operator one has to
choose again properly the boundary conditions which couple the
wavefunctions at the branching points of the graph in such a way
that the probability current conservation at the vertices is conserved.
As we have said, we shall require continuity of the line component
$\,f\,$ of $\,\psi\,$ at the junctions; to keep a manageable number of
parameters we assume the Dirichlet conditions at the end of the stubs.
Consequently, the domain of the Hamiltonian consists of all
$\,\psi\in\HH\,$ with $\,f\in AC^2(\R)\,$ and $\, u_s\in
AC^2(0,L_s)\,$ satisfying the conditions \cite{ESe}
\begin{eqnarray} \label{comb bc}
f((s\!-\!1)\ell+) &\!=\!& f((s\!-\!1)\ell-) \,=:\, f((s\!-\!1)\ell)\,,
\nonumber \\
u_s(0) &\!=\!& b_sf((s\!-\!1)\ell)+c_su'_s(0) \,, \nonumber \\
f'((s\!-\!1)\ell+)-f'((s\!-\!1)\ell-) &\!=\!& d_sf((s\!-\!1)\ell)
-b_su'_s(0)\,, \\ u_s(L_s) &\!=\!& 0  \nonumber
\end{eqnarray}
for $\,s=1,\dots,N\,$. In general, $\,b_s\,$ may be complex, however,
we restrict from the start to Hamiltonians which are time--reversal
invariant and suppose that the coefficient matrices $\,\K_s= \left(
\matrix{ b_s & c_s \cr d_s & -b_s} \right)\,$ are real. The $\,s$--th
appendix is decoupled from the line by putting $\,b_s=0\,$; it is
then described by the operator $\,h_{c_s}:= -\,{d^2\over dx^2}
\,+V_s\,$ specified by the decoupled condition
\begin{equation} \label{appendix bc}
u_s(0)-c_su'_s(0)\,=\,0
\end{equation}
at the junction.

In what follows we concentrate on the finite periodic case where
$\,L_s=L\,,\; V_s=V\,$, and $\,\K_s=\K\,$; the operator
specified by the boundary conditions (\ref{comb bc}) will be denoted
as $\,H_N\equiv H_N(\K,V)\,$. The potential $\,V\,$ is supposed to
belong to $\,L^1(0,L)\,$.


\subsection{Scattering on a comb graph}

To write the scattering matrix we need some notation. Let $\,u_L$ be
the unique solution to the appendix Schr\"odinger equation,
\begin{equation} \label{tooth}
-u_L''+Vu_L\,=\, k^2 u_L
\end{equation}
with the energy $\,k^2$ and the normalized Dirichlet condition at the
outer endpoint, $\,u_L(L)= 1\!-\!u'_L(L)=0\,$. If all the
junctions are described by the parameters $\,b,c,d\,$ in (\ref{comb bc}),
we put
\begin{equation} \label{beta z}
\beta\,:=\, {d\over 2k}\,+\,{b^2\over 2k}\, \left(
u'_L\over cu'_L\!-\!u_L \right)(0)\,, \qquad \zeta\,:=\, \cos
k\ell+\beta\sin k\ell\,,
\end{equation}
and denote again $\,\eps:= e^{ik\ell}$. For a single stub, $\,s=N=1\,$,
we insert a generalized eigenvector of the form $\, f \choose
\alpha u_1 \,$ into the boundary conditions (\ref{comb bc}) and
express $\,f(0+),\, f'(0+)\,$ by means of the left--sided limits; this
yields the single--element transfer matrix which can be with the help
of (\ref{beta z}) written as
$$
L_\beta\,=\, \left( \begin{array}{cc} 1 & 0 \\ 2k\beta & 1 \end{array}
\right)\,,
$$
so $\,R=1\!+\!i\beta\,$ and $\,S=i\beta\,$. Furthermore,
$\,Re(\overline\eps R)=\zeta\,$. Using then the explicit expression
(\ref{serial phase}) of the phase factor together with (\ref{t_N})
and (\ref{r_N}), we may write the left-to--right scattering amplitudes
as
\begin{equation} \label{comb rt_N}
r_N\,=\, {i\beta U_{N-1}(\zeta)\over \eps U_{N-2}(\zeta)-
(1\!+\!i\beta) U_{N-1}(\zeta)}\,, \quad t_N\,=\,-\, {\eps^{1-N}
\over \eps U_{N-2}(\zeta)- (1\!+\!i\beta) U_{N-1}(\zeta)}\,.
\end{equation}
In particular, the reflection and transmission probabilities are
\begin{equation} \label{comb rt_N^2}
|t_N(k)|^2\,=\, 1-|r_N(k)|^2\,=\, \frac{1}{1+\beta^2
U_{N-1}(\zeta)^2}\,.
\end{equation}
In order to illustrate the usefulness of the formulas
(\ref{t_N})--(\ref{t_N^2}), we present in Appendix A two other ways
to derive the relations (\ref{comb rt_N}).

Let us now illustrate how the the transmission probability depends on
the number of the teeth and the parameters of the junctions.
Fig. 4 shows a typical example of the situation. For a certain set of parameters
$b, c,$ and $d$ we compare $|t(k)|^2$ for one and seven appendices with
an infinite array of them. The band spectrum is marked by thick lines above.
The suppression of transmission coefficient at those values of $k$ where the
infinite system shows gaps is evident even for a relatively small number of
elementary scatterers, $N = 7$.
In each band we distinguish $N-1$, \ie \, six values of $k_i$ when $|t(k_i)|^2 = 1$
as it should be -- \cf (\ref{band Chebyshev}). In addition the first and third
band contain an additional value of $k_a$ for which $|t(k_a)|^2 = 1$ and these
values coincide with the values of $k$ for which a single appendix allows a
perfect transmission. The values of $k$ for which the reflection is total fall
into gaps of the infinite system as it should be (\cf the end of sec. 2.4).


\subsection{The resolvent poles}

The energy dependence of the transmission probability through the comb
graph is related to its resonance structure which in turn comes from
perturbation of embedded eigenvalues of $\,H_N(0,V)\,$ referring to
the stubs by the coupling between the line and the ``teeth". The
resonances are conventionally associated with the poles of the
analytically continued resolvent; we are going to show now how they
can be found.

As in the particular case $\,N=1\,$ discussed in \cite{ESe}, one can
derive an expression for the resolvent $\,(H_N\!-z)^{-1}$  of the
operator $\,H_N(\K,V)\,$ by means of Krein's formula
\cite[Appendix~A]{AGHH}: it is an integral operator with the kernel
\begin{equation} \label{Krein}
(H_N\!-z)^{-1}(x,y) = (H_0\!-z)^{-1}(x,y)+\sum_{m,n=1}^{2N}
\lambda_{mn} F_m(x,z) F_n(y,z)\,;
\end{equation}
in this relation $\,H_0\,$ is a suitable comparison operator and
the vectors $\,F_m\,$ satisfying $\,\overline{F_m(x,\bar z)}=
F_m(z)\,$ belong to the the deficiency subspace (for a given complex
energy $\,z\,$) of the maximum common restriction of $\,H_N\,$ and
$\,H_0\,$.

As usual we put $\,k:=\sqrt{z}\,$. We choose for $\,H_0\,$ the fully
decoupled operator with $\,b_s=c_s=d_s=0\,,\; s=1,\dots,N\,$. Its
resolvent can be written down explicitly; its kernel is a
$\,(N\!+\!1)\times (N\!+\!1)\,$ matrix
\begin{equation} \label{free kernel}
(H_0\!-z)^{-1}(x,y)\,=\, \left(\begin{array}{cccc}
R_1(x,y;z) & 0 & \ldots & 0 \\
0  & -\frac{u_{0,1}(x_<)u_{L,1}(x_>)}{W(u_{0,1},u_{L,1})} & \ldots & 0 \\
\vdots &  & \ddots & \vdots \\
0 &  &  & -\frac{u_{0,N}(x_<)u_{L,N}(x_>)}{W(u_{0,N},u_{L,N})}
\end{array} \right)\,,
\end{equation}
where $\,W(u_{0,n},u_{L,n})\,$ is the Wronskian of the solutions
$\,u_{0,n}\,$ and $\,u_{L,n}\,$ to $\,-u''+ V_n u = k^2u\,$
specified by the boundary conditions $\,u_{0,n}(0) = 0\,$ and
$\,u_{L,n}(L_n) = 0\,$, respectively. The symbols $\,x_<\,$ and
$\,x_>\,$ mean respectively the smaller and larger of the variables
$\,x,y\,$, and
\begin{equation} \label{line kernel}
R_1(x,y;z) \,:=\, \frac{i}{2k}e^{ik|x-y|}
\end{equation}
is the free resolvent kernel on the line. A natural choice of the
deficiency vectors is the following:
\begin{equation} \label{def vectors}
F_n(x) \,:=\, \left(\begin{array}{c}
           R_1(x,(n\!-\!1)\ell;z) \\ 0 \\ \vdots \\ 0
           \end{array}\right)\,, \qquad
F_{n+N}(x) := \left(\begin{array}{c}
           0 \\ \vdots \\ u_{L,n}(x,z) \\ \vdots \\ 0
           \end{array}\right)
\end{equation}
for $\,n=1,\dots,N\,$. To find the coefficients $\,\lambda_{mn}\,$
we use the fact that for $\,z\,$ of the resolvent set the resolvent
maps $\,L^2(\Gamma_N)\,$ onto the domain of $\,H_N\,$. Hence the
components of the vector at the \lhs of
\begin{equation} \label{f=R_H g}
\left(\begin{array}{c}
f \\ u_1 \\ \vdots \\ u_N
\end{array} \right)=(H_N\!-z)^{-1}
\left(\begin{array}{c}
g \\ v_1 \\ \vdots \\ v_N \end{array}\right)
\end{equation}
have to satisfy the boundary conditions (\ref{comb bc}) for arbitrary
(square integrable) functions $\,g,v_1,\dots,v_N\,$. For the sake of
brevity we introduce
$$
h_n := \int_{\R} R_1(y,(n\!-\!1)L;z)g(y)\,dy \,, \qquad
h_{n+N} := \int_0^{\ell_n}u_n(y,z)v_n(y)\,dy
$$
with $\,n=1,\dots,N\,$; they represent $\,2N\,$ independent
quantities. By direct computation we find then the needed boundary
values from (\ref{Krein})--(\ref{f=R_H g}),
\begin{eqnarray}\label{comb boundary values)}
f((s\!-\!1)\ell) &\!=\!& h_s +
\frac{i}{2k}\, \sum_{l=1}^{2N}\left(
\sum_{n=1}^N\eps^{|s-n|}\lambda_{nl}\right)h_\ell\,,\nonumber\\
f'((s\!-\!1)\ell+) &\!=\!& \int_{\R} \left. \frac{\partial R_1(x,y;z)}
{\partial x}\right|_{x=(s-1)\ell+}\, g(y)\,dy + \nonumber \\
&& + \frac{1}{2}\,\sum_{l=1}^{2N} h_l\left(-\sum_{n=1}^{s-2}
\lambda_{nl} \eps^{s-n} +
\sum_{n=s-1}^N\lambda_{nl}\eps^{n-s}\right)\,. \\
f'((s\!-\!1)\ell-) &\!=\!& \int_{\R} \left. \frac{\partial R_1(x,y;z)}
{\partial x}\right|_{x=(s-1)\ell-}\, g(y)\,dy + \nonumber \\
&& + \frac{1}{2}\,\sum_{l=1}^{2N} h_l \left(-\sum_{n=1}^{s-1}
\lambda_{nl} \eps^{s-n}+ \sum_{n=s}^N
\lambda_{nl}\eps^{n-s}\right)\,,  \nonumber \\
u_s(0) &\!=\!& u_{L,s}(0) \sum_{k=1}^{2N} h_k\lambda_{N+s,k}\,,
\nonumber \\
u_s'(0) &\!=\!& \frac{1}{u_{L,s}(0)}\, h_{N+s} +
u_{L,s}'(0)\,\sum_{k=1}^{2N} h_k\lambda_{N+s,k} \,, \nonumber
\end{eqnarray}
where $\,s=1,\dots,N\,$ and $\,\eps:=e^{ik\ell}\,$. Substituting from
here to (\ref{comb bc}) and using the fact that $\,h_n\,$ are independent
we arrive at the following system of equations
\begin{eqnarray} \label{coef system}
\sum_{n=1}^N\, \frac{ib_s}{2k}\,\eps^{|s-n|} \lambda_{nl} \,+\,
\left(c_s u_{L,s}'(0)- u_{L,s}(0)\right)\lambda_{N+s,l} &\!=\!&
\left\{\begin{array}{ll}
-b_s & \quad l=s \\
-\,\frac{c_s}{u_{L,s}(0)} & \quad l=N\!+\!s \\
\phantom{A} 0 & \quad \mbox{otherwise} \end{array}\right.
\nonumber \\ \\
\sum_{n=1}^N\, \left(\frac{id_s}{2k}\,\eps^{|s-n|}
\,+\,\delta_{sn}\right)\lambda_{nl} \,-\,b_s
u_{L,s}'(0)\lambda_{N+s,l} &\!=\!& \left\{
\begin{array}{ll} -d_s & \quad l=s \\
\frac{b_s}{u_{L,s}(0)} & \quad l= N\!+\!s \\
\phantom{A} 0 & \quad \mbox{otherwise} \end{array}\right. \nonumber
\end{eqnarray}
with $\,s=1,\dots,N,\; l=1,\dots,2N\,$. If we arrange the sought
coefficients in the following way,
$$
\lambda_{1,1}, \dots,\lambda_{2N,1}, \lambda_{1,2}, \dots,
\lambda_{2N,2}, \;\dots,\; \lambda_{1,2N}, \dots,\lambda_{2N,2N}\,,
$$
the matrix of the system (\ref{coef system}) is block diagonal with
$\,2N\,$ same $\,2N\times 2N\,$ blocks. This is important because we
are looking for singularities of the resolvent (\ref{Krein}) which
occur at the points where the coefficients $\,\lambda_{mn}\,$ are
singular. The latter are solutions to the linear system (\ref{coef
system}), and therefore they have the same denominator; looking for
its zeros it is sufficient to inspect the determinant of a single
block,
$$
\left| \begin{array}{ccc} {ib_s\over 2k}\,\eps^{|s-n|} & \vdots &
\left(c_s u_{L,s}'- u_{L,s}\right)(0)\, \delta_{s,N-n} \\
\dots & \dots & \dots \\ \frac{id_s}{2k}\,\eps^{|s-n|}
\,+\,\delta_{s-N,n} & \vdots & -\,b_{s-N} u_{L,s-N}'(0) \delta_{s,n}
\end{array} \right|\,.
$$
For a non--real $\,z\,$ we have
$$
\left(c_s u_{L,s}'- u_{L,s}\right)(0)\ne 0 \qquad {\rm and} \qquad
\,b_{s} u_{L,s}'(0)\ne 0
$$
for all $\,s=1,\dots,N\,$, since otherwise the decoupled stub
Hamiltonian with the appropriate boundary condition at $\,x=0\,$
would have a complex eigenvalue which is clearly impossible.
Vanishing of the above determinant is then equivalent to
$$
D_N\,:=\, \det\left( i\beta_n\eps^{|s-n|}+ \delta_{s,n} \right)
\,=\,0\,,
$$
where
$$
\beta_n\,:=\, {d_n\over 2k}\,+\,{b_n^2\over 2k}\, \left(
u'_{L,n}\over cu'_{L,n}\!-\!u_{L,n} \right)(0)\,.
$$
Up to now the stubs could be mutually different. We shall evaluate
the determinant in the finitely periodic case, $\,\beta_n=\beta\,$
for $\,n=1,\dots,N\,$. Then $\,D_N\,$ can be written as
\begin{equation} \label{differ eq}
D_N\,=\, {1\over 1\!-\!\eps^2}\, \left\lbrack (i\beta(1\!-\!\eps^2)
+1)\tilde D_{N-1} - \eps^2\tilde D_{N-2} \right\rbrack\,,
\end{equation}
where
\begin{equation} \label{init cond}
D_0\,=\, i\beta+1\,, \qquad D_1\,=\, 1+2i\beta-\beta^2
(1\!-\!\eps^2)\,,
\end{equation}
and the quantities satisfy the recursive relation
\begin{equation}
\overline\eps \tilde D_N -2z\tilde D_{N-1}
+\eps\tilde D_{N-2}\,=\, 0\,.
\end{equation}
Comparing this to the relations between Chebyshev polynomials
we find
\begin{equation}
D_N\,=\, -\eps^{N-1} \left( \eps U_{N-2}(\zeta)- (1\!+\!i\beta)
U_{N-1}(\zeta) \right)\,,
\end{equation}
since (\ref{differ eq}) is a linear difference equation of the
second order with constant coefficients which has for the given
initial conditions (\ref{init cond}) a unique solution.

We see that $\,D_N(k)=0\,$ if and only if the same is true for
the denominator of the reflection and transmission amplitude. Hence
the poles of the continued resolvent and of the S--matrix generically
coincide. It might be that some of them vanish due to a zero in the
numerator, however, this can happen in isolated points only because
the numerators are analytical functions of the coupling parameters,
and there is a one--to--one correspondence between the pole
trajectories with respect to the coupling constant and the embedded
eigenvalues of the decoupled Hamiltonian $\,H(0,V)\,$.


\subsection{Band spectrum of an infinite comb and resonances}

We discuss only the case of free appendices here, \ie $\,V_s = 0$.
Investigation of non-zero potentials represents no complication in
general conception. Typical spectra of an infinite comb are
presented on Fig. 5. The parameters of such a structure could be
divided into two groups, \viz $\ell,L$ and $b, c, d$. The dependence
on $\ell$ is rather simple; a change of $\ell$ results in scaling of
spectrum in $k$. This is the consequence of the fact that the transfer
matrix $T$ depends only on the product $k\ell$. This is the reason why
we show all the spectra for $\ell = 1$ only. The dependence on $L$ is
more complicated as can be inferred from Fig. 5.

A striking feature of such a spectrum are sudden transitions of bands
to gaps or vice versa at $k\ell = n\pi,\,\, n\in{\Z}$ for almost all $L$.
This can be understood, if we write down the Bloch condition explicitly
as \[\beta(k) \sin(k\ell) + \cos(k\ell) = \cos(\theta).\] Clearly,
this is always fulfilled for $k\ell = n\pi$, unless $\beta$ has a
singularity at $n\pi/\ell$. Therefore $k = n\pi/\ell$ belongs always to
the band in this case. The situations when $\beta(n\pi/\ell)$ is infinite
requires a further analysis.
An effect of a small change $\epsilon$ in $k$ can be  estimated from
$\beta(k+\epsilon) \epsilon\ell + 1 = (-1)^n \cos(\theta)$.
With exception of points where $\beta(n\pi/\ell) = 0$, small change
of $\epsilon$ does not change the sign of $\beta$, while the change of
sign of $\epsilon$ leads to the change of sign of
$\beta(k+\epsilon) \epsilon\ell$. Only at vicinity of points where
$\beta(n\pi/\ell) = 0$ can the sign of $\beta(k+\epsilon) \epsilon\ell$
rest unchanged, \cf Fig 5.

Of the three parameters determining the coupling of an appendix
to the backbone line \ie $\,\,b, c, d$, we find $b$ as the most suitable
to begin with. Putting $b = 0$ switches effectively off the appendices
as already mentioned in sec. 3.2. This implies that we have eigenvalues
$k_n^2$ embedded in the continuum, where $k_n$ are solutions of
(\ref{appendix bc}).
These values of momentum play an important role from three points of view.
First, $k_n$'s are the values for which $u_L(0) = 0$ and the number of
zeros of $u$ increases by one when $k$ passes to higher values; this
means that the whole ${\R}^+$ is divided into disjoint intervals by
these values (for a given $L$). On the other hand, a band can belong to
two intervals (although this happens only for some $L$) and states of
the same band can be described by wave functions with different number
of zeros of $u$'s. Also, and this situation is far more frequent, one
interval can contain two or more band, so that the number of zeros of
$u$ is not directly related to the ordinal number of a band.

In order to reveal the second role of these values $k_n$'s, we have to
return to finite number of appendices attached to the line. If we begin
with one appendix only, direct calculation shows that $t(k_n) = 0$ and
$r(k_n) = -1$. Therefore, regardless of the number of appendices $N$
the full reflection takes place, \ie $\,\,t_N = 0, r_N = -1$.

We close this section with discussion of scattering resonances of our
system. Recently, it was shown \cite{BG} that there is certain ``band"
structure in the spectrum of resonances and that this spectrum converges
to the energy band spectrum of the infinite periodic system in the limit
of an infinite number of scatterers. The authors also show that one
should expect $N - 1$ resonances in each band if the system consists of
$N$ identical cells. These conclusions are in accordance with our results
but for one difference -- the number of resonances in a band is $2N - 1$.
It is a consequence of the fact that the origin of resonances is twofold
here. $N - 1$ resonances comes from the spacial order of our periodical
structure as in \cite{BG} and the remaining $N$ ones have their origin
in $k_n^2$, which is an $N$-fold eigenvalue for the case of $b = 0$. As
$|b|$ grows this degenercy is lifted and we find $N$ resonances in the
vicinity of $k_n^2$. This is the third role of $k_n$. The $N - 1$
resonances related to the spatial setting travel from $-\infty$ and for
small $b$ these two sets are well separated as is shown on Fig. 6.

There are situations when a resonance or a bound state appears at $k = 0$.
The manifestation of this instance is that $|t(0)|^2 = 1$. This can take
place even in the case of one appendix. The expression (5) of \cite{ESe}
with explicit form of $u_L(0)$ shows that this can happen as soon as the
condition \[d(c + L) + b^2 = 0\] is fulfilled. Provided that this condition
holds, there is perfect transmission at $k = 0$ for any number of appendices,
\ie $\,t_N(0) = 1$. Besides this there could be other situations when
$t_N(0) = 1$ for $N > 1$. If there exists such an $n = 1,\ldots,N-1$ that
\[ \cos\left(\frac{n\pi}{N}\right) = 1 + \frac{(d c + d L + b^2)\ell}{2(c + L)}, \]
we have again $t_N(0) = 1$. This corresponds to $U_{N-1}(\zeta(0)) = 0$ in
(\ref{comb rt_N^2}). Direct inspection confirms that the wave function does
not belong to $L^2(\Gamma_N)$ and that these states are resonances (and not
bound states).


\setcounter{equation}{0}
\section{Serial structures of mixed dimensionality}

In this section we want to treat the situation when the scatterers
connected by single--mode leads have a higher dimension. For the
sake of simplicity, we shall consider only the simplest
possibility when the dimension is two, \ie, the scatterer is a
surface. Such systems can be realized in both the solid state
(recall, \eg, the ``bamboo defects" in nanotubes \cite{KBR}) and
electromagnetism (for flat resonators), but we avoid discussing
examples and the conditions under which these models are
realistic. We suppose that the surface is smooth, bounded, and
connected, with or without the boundary.  Although it makes no
difficulty to let the  particle on the surface  interact with an
external potential field, we  will regard it as free, \ie, its
Hamiltonian will be (in appropriate units) just the corresponding
Laplace--Beltrami operator.


\subsection{Coupling of leads to a surface}

The basic question for the described serial structures is the way
in which the leads are coupled to the scatterers. The physical condition
is again a conservation of the probability current, which translates into
the self--adjointness requirement of the corresponding Hamiltonian. Since
the coupling is local, we may disregard geometrical peculiarities of
the lead and the surface and consider the setting when a halfline is
attached to a plane. The state Hilbert space is then $\,L^2({\R}^-)
\oplus L^2({\R}^2)\,$ and the Hamiltonian acts on its elements $\,\phi_1
\choose \phi_2\,$ as $\,-\phi''_1 \choose -\Delta\phi_2\,$. To make it
self--adjoint one has to impose suitable boundary conditions which
couple the wavefunctions at the junction. A general solution to this
problem is given in Ref.\cite{ES1}. The conditions are of the form
\begin{equation} \label{12 bc}
\phi_1^{'}(0-) \,=\, A\phi_1(0-)+BL_0(\phi_2)\,, \qquad
L_1(\phi_2) \,=\, C\phi_2(0-)+DL_0(\phi_2)
\end{equation}
together with several ``exceptional" classes, where
\begin{eqnarray*}
L_0(\phi_2) &\!=\!& \lim_{r\to 0+}\frac{\phi_2(\vec x)}{\ln r}\,,
\\ \\ L_1(\phi_2) &\!=\!&
\lim_{r\to 0+}\left(\phi_2(\vec x)-L_0(\phi_2)\ln r\right)
\end{eqnarray*}
with $\,r:=|\vec x|\,$ are the generalized boundary values in the plane,
and the coefficients $\,A,B,C,D\,$ depend on four real parameters;
evaluating the boundary  form, it is straightforward to see that they
satisfy restrictions
\begin{equation} \label{12 coef cond}
A, D\in {\R}\, \qquad B = 2\pi\overline C\,.
\end{equation}
A disadvantage of this result is that it tells us nothing about
physical relevance of the coefficients values in the boundary conditions
(\ref{12 bc}). The choice  of the coupling depends on particular
properties of the junction it models, of course, but one would like to
select a subclass representing a ``natural" coupling.  One way to achieve
this goal was suggested  in \cite{ES3}:  comparing the scattering matrix
of the junction given by (\ref{12 bc}) with the  low--energy behavior of
scattering in the system  of a plane to which  a cylindrical ``tube" is
attached, and taking into account the condition (\ref{12 coef cond}), we
arrive at  the identification
\begin{equation} \label{12 phys coupling}
A \,=\, {1\over 2\rho}\,,\qquad B = \sqrt{{2\pi\over \rho}}\,, \qquad
C = {1\over\sqrt{2\pi\rho}}\,, \qquad D = -\ln\rho\,,
\end{equation}
where $\,\rho\,$ is the contact radius. Physical relevance of these
conditions was illustrated in \cite{ES3} by explaining the experimentally
observed distribution of resonances in a microwave resonator with a
thin antenna. Motivated by this, we will use in the following
(\ref{12 bc}) and (\ref{12 phys coupling}) to describe the coupling
between the leads and the scatterers.


\subsection{The single--element S--matrix}

Using the local character of the boundary conditions derived above we
apply them to coupling of a pair of halfline leads to an arbitrary
surface $\,G\,$. The  only restriction is that that the junction may
not belong to the boundary of $\,G\,$ if it has any. We shall compute
the transfer and scattering matrices for such a system.

As we have said the Hamiltonian is a Laplace--Beltrami operator on the
state Hilbert space $\,L^2(G)\,$ of the scatterer. We shall characterize
it by its Green's function $\,G(.,.;k)\,$, \ie, the integral kernel of
its resolvent which exists whenever $\,k^2$ does not belong to the
spectrum. Its actual form depends on the geometry of $\,G\,$ but we shall
not need it. What is important is the character of its singularity. As
a smooth manifold, $\,G\,$ admits in the vicinity of any point a local
Cartesian chart and the Green's function behaves as that of Laplacian
in the plane,
\begin{equation}
G(x,y;k)\,=\,-\,\frac{1}{2\pi}\,\ln |x\!-\!y| +\OO(1)\,,
\qquad |x\!-\!y|\to 0\,.
\end{equation}
Looking for scattering solutions to the Schr\"odinger equation, we need
a general solution to the Laplace--Beltrami equation on $\,G\,$ for the
energy $\,k^2$. Without loss of generality, we may write it as
\begin{equation} \label{surface soln}
u(x) \,=\, a_1G(x,x_1;k)+a_2G(x,x_2;k)\,,
\end{equation}
where $\,x_1,x_2\,$ are two different points of $\,G\,$ at which the
leads are attached. The generalized boundary values (labelled by the
point at which they are taken) of this solution are then
\begin{equation} \label{gen bv}
L_0[x_j] \,=\, -\frac{a_j}{2\pi}\,,\qquad L_1[x_j] \,=\,
a_j\xi(x_j,k)+a_{3-j}G(x_1,x_2;k)
\end{equation}
for $\,j=1,2\,$, where
\begin{equation} \label{xi1}
\xi(x_j;k) \,=\, \lim_{x\to x_j} \left\lbrack G(x,x_j;k)+
\frac{\ln|x\!-\!x_j|}{2\pi} \right\rbrack\,.
\end{equation}

Next we denote the wavefunction on the $\,j$--th lead as $\,u_j\,$.
For simplicity we use the abbreviations $\,u_j, \,u'_j\,$ for its
boundary values; then the boundary conditions (\ref{12 bc}) yield
\begin{eqnarray} 
u'_1\,=\, A_1u_1-\,\frac{B_1a_1}{2\pi} \,, && \qquad
a_1\xi_1+a_2g \,=\, C_1u_1-\,\frac{D_1a_1}{2\pi} \,, \nonumber \\
\nonumber \\
u'_2\,=\, -A_2u_2+\,\frac{B_2a_2}{2\pi} \,, && \qquad
a_2\xi_2+a_1g \,=\, C_2u_2-\,\frac{D_2a_2}{2\pi} \,, \nonumber
\end{eqnarray}
where $\,g:=G(x_1,x_2;k)\,$.
In the the first equation of the second pair we have changed sign,
because the second lead is identified with $\,{\R}^+$. It is
straightforward to rewrite these equations as a linear system with
the unknown $\,u_2,\, u'_2,\, a_1,\, a_2\,$ and to solve it; this gives
in particular the transfer matrix,
\begin{equation} \label{geom L}
L \,=\, \frac{1}{gC_2}\left(\begin{array}{cc}
C_1Z_2\!+\!2\pi\frac{A_1}{B_1}\Delta & -2\pi\frac{\Delta}{B_1}
\\ \\
B_2C_2\left( \frac{C_1}{2\pi}\!-\!Z_1 \frac{A_1}{B_1} \right) -
C_1A_2Z_2-2\pi\frac{A_1A_2}{B_1}\Delta &
2\pi\frac{A_2}{B_1}\Delta + \frac{B_2C_2Z_1}{B_1}
\end{array}\right)\,,
\end{equation}
where $\,Z_j:= {D_j\over 2\pi} +\xi_j\,$ and $\,\Delta :=
g^2\!-Z_1Z_2\,$. Using (\ref{12 coef cond}) we  find easily
\begin{equation}
\det L \,=\, -\frac{B_2C_1}{B_1C_2} \,=\,
-\,\frac{\overline C_2C_1}{\overline C_1C_2}\,,
\end{equation}
so $\,\det L=1\,$ if the junctions are identical or the coefficients
$\,C_j$ are real. The second possibility if the couplings are invariant
with respect to the time reflection, which we shall suppose in the
following. In that case the transfer matrix simplifies to the form
\begin{equation}  \label{geom L2}
L \,=\, \frac{1}{g}\left(\begin{array}{cc}
Z_2+\frac{A}{C^2}\Delta & -2\frac{\Delta}{C^2} \\ \\
C^2-A(Z_1\!+\!Z_2)-\frac{A^2}{C^2}\Delta &
\frac{A}{C^2}\Delta +Z_1
\end{array}\right)\,,
\end{equation}
in particular
\begin{equation} \label{geom L3}
L \,=\, \frac{1}{g}\left(\begin{array}{cc}
Z_1\!+\!\pi\Delta & -2\pi\rho\Delta \\ \\
\frac{1}{2\rho}\left(\frac{1}{\pi}-Z_1\!-\!Z_2\!
-\!\pi\Delta\right) & Z_1 \!+\!\pi\Delta
\end{array}\right)
\end{equation}
for the physically most interesting class of couplings (\ref{12 phys
coupling}). The S--matrix of our geometric scatterer is then given by
the relations (\ref{t_N}) and (\ref{r_N}); in the case (\ref{geom L3})
we have
\begin{eqnarray}  \label{geom S}
r(k) \,=\,
-\,\frac{\pi\Delta+Z_1+Z_2-\pi^{-1}+2ik\rho(Z_2\!-\!Z_1)
+4\pi k^2\rho^2\Delta}{\pi\Delta+Z_1\!+Z_2\!-\pi^{-1}
+2ik\rho(Z_1\!+\!Z_2\!+\!2\pi\Delta) -4\pi k^2\rho^2\Delta}\,,
\nonumber \\ \\
t(k) \,=\, -\frac{4ik\rho g} {\pi\Delta+Z_1\!+Z_2\!-\pi^{-1}
+2ik\rho(Z_1\!+\!Z_2\!+\!2\pi\Delta) -4\pi k^2\rho^2\Delta}\,.
\nonumber
\end{eqnarray}

To make use of these formulas, we need to know $\,g,\, Z_1,\, Z_2,\,
\Delta\,$ as functions of the momentum $\,k\,$. By assumption the
manifold $\,G\,$ is compact, so the spectrum $\,\{\lambda_n\}_{n=1}
^{\infty}\,$ of the Hamitonian is purely discrete and the corresponding
eigenfunctions $\,\{\phi(x)_n\}_{n=1}^{\infty}\,$ form an orthonormal
basis in $\,L^2(G)\,$. The usual Green's function expression then gives
\begin{equation} \label{g}
g(k)\,=\,\sum_{n=1}^{\infty}\, \frac{\phi_n(x_1)
\overline{\phi_n(x_2)}}{\lambda_n\!-k^2}\,.
\end{equation}
To express the remaining three values we have to compute the regularized
limit (\ref{xi1}). Expanding the logarithm into the Taylor series, we
can rewrite the sublimit expression as
$$
G(x_j+\sqrt{\eps}n,x_j;k)+\frac{\ln{\sqrt\eps}}{2 \pi}
\,=\, \sum_{n=1}^\infty\left(\frac{\phi_n(x_j+\sqrt\eps
n)\phi_n(x_j)}{\lambda_n-k^2}-\frac{(1\!-\!\eps)^n}{4\pi
n}\right)\,,
$$
where $\,n\,$ is a unit vector in the local chart around the point
$\,x_j\,$. Unfortunately, interchanging the limit with the sum is not
without risk since the latter does not converge uniformly. To see that
the result may indeed depend on the regularization procedure, it is
sufficient to replace $\,\sqrt\eps\,$ by $\,c\sqrt\eps\,$ at
the \lhs. To make idea about this non--uniqueness, let us compute the
difference
$$
\xi(x_j,k) - \xi(x_j,k') \,=\, \lim_{\eps\to
0+}\sum_{n=1}^{\infty}\left(\frac{\phi_n(x_j+\sqrt{\eps
n})\overline{\phi_n(x_j)}}{\lambda_n\!-k^2} \,-\,\frac{\phi_n(x_j
+\sqrt{\eps n})\overline{\phi_n(x_j)}}{\lambda_n-k'}\right)\,.
$$
This sum is already uniformly convergent, because by standard
semiclassical estimates \cite[XIII.16]{RS} the sequence $\,\{
\|\phi_n\|_\infty\}_{n=1}^{\infty}\,$ is bounded with our assumptions
and $\,\lambda_n= 4\pi|G|^{-1}n +\OO(1)\,$ as $\,n\to\infty\,$, so
$$
\frac{1}{\lambda_n\!-k^2}\,-\,\frac{1}{\lambda_n\!-k^{'2}}
\,\sim\,\frac{1}{n^2}\,,
$$
and therefore
\begin{equation} \label{diff1}
\xi(x_j,k) - \xi(x_j,k') \,=\,
\sum_{n=1}^{\infty}\,\left(\frac{|\phi_n(x_j)|^2}{\lambda_n\!-k^2}
\,-\,\frac{|\phi_n(x_j)|^2}{\lambda_n\!-{k'}^2}\right)\,.
\end{equation}
From the same reason
\begin{equation}
\tilde\xi (x_j,k)\,: =\,
\sum_{n=1}^{\infty}\,\left(\frac{|\phi_n(x_j)|^2}
{\lambda_n\!-k^2}\,-\,\frac{1}{4\pi n}\right)
\end{equation}
makes sense and $\,\xi (x_j,k)- \tilde\xi(x_j,k)\,$ is independent of
$\,k\,$. We have therefore
\begin{equation} \label{xi2}
\xi (x_j,k) \,=\, \sum_{n=1}^{\infty}\, \left(\frac{|\phi_n(x_j)|^2}
{\lambda_n\!-k^2}\,-\,\frac{1}{4\pi n}\right) \,+\, c(G)\,.
\end{equation}
The constant depends only on the manifold $\,G\,$ we will neglect it
in the following, because its nonzero value means just a coupling
constant renormalization: $\,D_j\,$ has to be changed to $\,D_j\!
+2\pi c(G)\,$. For a flat rectangular $\,G\,$ we found in \cite{ES3} an
agreement with the experiment using $\,c(G)=0\,$.


\subsection{A ``bubble" on the line}

To make the above consideration more concrete, we shall concentrate in
the rest of this section on a single example. We are going to consider
the case when $\,G\,$ is a sphere of radius $\,R\,$ with the leads
attached at the poles, which is the system proposed by Kiselev
\cite{Ki}. The most important result of this paper was that apart of
the resonances coming from the bound states on the sphere, such a
scatterer has the high--energy behavior similar to that of the
so--called $\,\delta'$ interaction \cite{ADE,AEL,E5}, \ie, the
transmission probability {\em decays} as $\,E^{-1}$ for $\,E\to\infty\,$.
This was established in Ref. \cite{Ki} up to a logarithmic correction.
The difference here is that we
shall use the physically interesting coupling (\ref{12 phys coupling}),
while \cite{Ki} employed a two--dimensional subset  of the conditions
(\ref{12 bc}) disjoint with the above one. This leads to a different
S--matrix, and while the indicated high--energy behavior remains preserved,
the argument used in \cite{Ki} to demonstrate it has to be changed in
numerous places; this is why we present its modified version here.

The sphere Hamiltonian is chosen in the standard way. Using the spherical
coordinates, we write it as
\begin{equation}
H_G\,=\,\frac{1}{R^2}
\frac{\partial^2}{\partial\theta^2}\,+\,
\frac{\cot\theta}{R^2}\frac{\partial}{\partial\theta}\,+\,
\frac{1}{R^2\sin^2\theta}\frac{\partial^2}{\partial\varphi^2}
\end{equation}
with the usual domain. For the sake of simplicity we put $\,R=1\,$ in
the following; the results for a general $\,R\,$ can be obtained by
a scaling transformation. The spectrum of $\,H_G\,$ then consists of
the eigenvalues are $\,\lambda_{l,m}=l(l\!+\!1)\,, \;l=0,1,\dots\,,\;
m=-l,-l\!+\!1,\dots,l\,$ of multiplicity $\,2l\!+\!1\,$ to which
the eigenfunctions
$$
\phi_l^m(\theta,\psi) \,=\,
\sqrt{\frac{(2l\!+\!1)(l\!-\!|m|)!}{4\pi(l\!+\!|m|)!}}
\,P_l^{|m|}(\cos\theta)\,e^{im\varphi}
$$
correspond. The junctions $\,x_1,\,x_2\,$ we place at the points
$\,\theta=0\,$ and $\,\theta=\pi\,$, respectively, where
$$
P_l^{|m|}(\pm1)\,=\,(-1)^l \delta_{0,|m|}\,,
$$
so only states with $\,m=0\,$ can be coupled to the leads. To express
the S--matrix, we need the quantities
\begin{eqnarray} \label{g bubble}
g(k)&\!=\!&\frac{1}{4\pi}\,\sum_{l=1}^\infty\,
\frac{2l\!+\!1}{l(l\!+\!1)-k^2}\,(-1)^l\,, \\ \nonumber  \\ \label{Z bubble}
Z_j(k) &\!=\!&Z(k)\,:=\,
\frac{1}{4\pi}\,\sum_{l=1}^\infty\,
\left(\frac{2l\!+\!1}{l(l\!+\!1)-k^2}\,-\,
\sum_{j=0}^{2l}\,\frac{1}{l^2\!+\!j\!+\!1}\right)
-\,\frac{\ln\rho}{2\pi}\,+\,c(G)\,, \phantom{AAA}
\end{eqnarray}
where $\,\rho\,$ is the junction diameter which is supposed to be the
same for $\,j=1,2\,$. The relations (\ref{geom S}) yield, in particular,
the transmission probability in the form
\begin{equation} \label{t_bubble}
t(k) \,=\, \frac{4\pi ik\rho g(k)}{1-2\pi
Z(k)(1\!+\!2ik\rho)-\pi^2\Delta(k)(1\!+\!2ik\rho)^2 }\,,
\end{equation}
where $\,\Delta(k)= (g(k)-Z(k))(g(k)+Z(k))\,$.

To prove that (\ref{t_bubble}) behaves as indicated above at large values
of $\,E=k^2$, we need several auxiliary results.
\begin{lemma}
The function $\,Z(\cdot)\!+\!(-1)^lg(\cdot)\,$ is strictly increasing
on the intervals $\,(l(l\!-\!1),l(l\!+\!1))\,$.
\end{lemma}
{\em Proof:} We have
$$
Z'(k)\!+\!(-1)^l g'(k)\,=\,2k\,\sum_{j=0}^{\infty} \frac{2j\!+\!1}
{(j(j\!+\!1)-E)^2}\, \left(1\!+\!(-1)^{j+l}\right)\,>\,0\,.
\quad \QED
$$
\begin{lemma}
We have
$$
g(k) \,=\, \frac{(-1)^l}{4\pi}\left(-\,\frac{2l\!-\!1}{l(l\!-\!1)-E}
\,+\,\frac{2l\!+\!1}{l(l\!+\!1)-E}\right)\,+\, \OO(1)
$$
for $\,E\in (l(l\!-\!1),l(l\!+\!1))\,$ and any positive integer $\,l\,$
with the error term independent of $\,l\,$. Moreover, there is a
$\,K>0\,$ such that $\,g(k)\ge K\,$ holds for all $\,k\,$ large enough.
\end{lemma}
{\em Proof:} The error can be estimated explicitly; we shall show that
\begin{equation} \label{g error}
\left|\,g(k)\,-\,\frac{(-1)^l}{4\pi}\left(-\,\frac{2l\!-\!1}{l(l\!-\!1)-E}
\,+\,\frac{2l\!+\!1}{l(l\!+\!1)-E}\right)\right| \,<\, \frac{1}{2\pi}\,.
\end{equation}
To this end we have to find a bound to (\ref{g bubble}) on the interval
$\,(l(l\!-\!1),l(l\!+\!1))\,$ with the two singular terms removed. The
terms to the left and to the right of this pair form alternative sequences
with the decreasing modulus, and as such each of them may be estimated by
the (modulus of) the first term of such a sequence, \ie, by
$$
\frac{1}{4\pi} \left|\frac{2l\!-\!3}{(l\!-\!1)(l-2)-E}\right|
\qquad {\rm and} \qquad
\frac{1}{4\pi} \left|\frac{2l\!+\!3}{(l\!+\!1)(l\!+\!2)-E}\right|\,,
$$
respectively. Taking the maxima of these expressions and summing
them we arrive at (\ref{g error}). To get the second claim, we have
have to compare this result with a lower bound to the modulus of the
sum of the two singular terms. The minimum $\,\pi^{-1}\!+ \OO(l^{-1})\,$
of the latter is reached at $\,E= l^2-\frac{1}{4}+ \OO(l^{-2})\,$,
so this part wins over the other one once the error terms become
small enough. \quad \QED
\begin{lemma}
For a positive integer $\,l\,$ and $\,E\in (l(l\!-\!1),l(\!l\!+1))\,$
we have
\begin{equation} \label{Z bubble2}
Z(k)\,=\,\frac{1}{4\pi}\left(\,\frac{2l\!-\!1}{l(l\!-\!1)-E} \,+\,
\frac{2l\!+\!1}{l(l\!+\!1)-E}\,\right)\,-\,\frac{\ln l}{2\pi}
\,+\,\OO(1)\,,
\end{equation}
where the error term is independent of $\,l\,$.
\end{lemma}
{\em Proof:} We again split the two singular terms in (\ref{Z bubble2})
and write the rest as $\,Z_-(k,l)+Z_+(k,l)- \,\frac{\ln\rho}{2\pi}\,
+c(G)\,$, where
\begin{eqnarray*}
Z_-(k,l) &\!:=\!& {1\over 4\pi}\,\sum_{j=1}^{l-2}
\left(\frac{2j+1}{j(j\!+\!1)-E}
\,-\,\sum_{n=0}^{2j}\,\frac{1}{j^2\!+\!n\!+\!1}\right)\,,\\ \\
Z_+(k,l) &\!:=\!& {1\over 4\pi}\,\sum_{j=l+1}^{\infty}
\left(\frac{2j+1}{j(j\!+\!1)-E}
\,-\,\sum_{n=0}^{2j}\,\frac{1}{j^2\!+\!n\!+\!1}\right)\,.
\end{eqnarray*}
It is easy to see that
$$
\sum_{n=0}^{2j}\,\frac{1}{j^2\!+\!n\!+\!1}
\,=\,\frac{2}{j}\,+\,\OO(j^{-2})
$$
as $\,j\to\infty\,$, so the $\,j$--th term in $\,Z_\pm(k,l)\,$ can be
estimated from above by
$$
\frac{1}{4\pi}\left(\frac{2j\!+\!1}
{j(j\!+\!1)-l(l\!+\!1)}\,-\,\frac{2}{j}\,+\,\OO(j^{-2})\right)
$$
and from below by the same expression with $\,l(l\!+\!1)\,$ replaced
by $\,l(l\!-\!1)\,$. Using the identity
$$
\frac{2j\!+\!1}{j(j\!+\!1)-l(l\!+\!1)}\,=\, {1\over j\!+\!l\!+\!1}
\,-\,{1\over l\!-\!j}
$$
we find
\begin{eqnarray*}
4\pi Z_-(k,l) &\!=\!&
\left(\,-\sum_{m=2}^{l-1}\,+\, \sum_{m=l+2}^{2l-1}\,
-2\sum_{m=1}^{l-2}\,\right) {1\over m} \,+\,\sum_{m=1}^{l-2}\, \OO(m^{-2})
\,=\, -3\ln l \,+\,\OO(1)\,, \\ \\
4\pi Z_+(k,l) &\!=\!&
\lim_{n\to\infty}\, \left\lbrack\left(\,\sum_{m=1}^{n-l}\,+\,
\sum_{m=2l+2}^{n+l+1}\, -2\sum_{m=l+1}^{n}\,\right){1\over m}
\,+\,\sum_{m=l+1}^{n}\, \OO(m^{-2}) \right\rbrack \\ \\
&\!=\!& \lim_{n\to\infty}\, \ln {(n\!-\!l)((n\!+\!l\!+\!1) \over n^2}
\,+\, \ln (l\!+\!1)\,+\,\OO(1) \,=\, \ln l\,+\,\OO(1)\,.
\end{eqnarray*}
Summing the expressions we get the upper bound in (\ref{Z bubble2});
the lower one is obtained in the same way. \quad \QED
\begin{lemma}
For any $\,l\,$ large enough the interval $\,(l(l\!-\!1),l(l\!+\!1))\,$
contains a point $\,\mu_l\,$ such that $\,\Delta(\sqrt{\mu_l})=0\,$. The
number $\,\mu_l\,$ has the following properties:
\begin{description}
\item{\em (i)} $\;l(l\!+\!1)-\mu_l = 2l(\ln l)^{-1}\left(1\!
+\!\OO(1)\right)\,$,
\item{\em (ii)} $\;\frac{2l\!+\!1}{l(l\!+\!1)-E}\,\leq\,\ln l\,+\OO(1)\,$
for $\,E\leq\mu_l\,$,
\item{\em (iii)} $\;\frac{2l\!-\!1}{l(l\!-\!1)-E}\,=\,\OO(1)\,$
for $\,E>\mu_l\,$,
\item{\em (iv)} finally, $\,\frac{Z(\sqrt{\mu_l})}{|g(\sqrt{\mu_l})|}
\,=\,-1+\OO((\ln l)^{-1})\,$.
\end{description}
\end{lemma}
{\em Proof:} Fix $\,l\,$. We have $Z(k)^2\!-g(k)^2=\,
\left(Z(k)\!+\!(-1)^lg(k)\right) \left(Z(k)\!-\!(-1)^lg(k)\right)\,$, and
by the preceding lemmas these expressions equal
$$
\frac{1}{2\pi}\, \frac{2l\!\pm\!1}{l(l\!\pm\!1)-E}
\,-\,\frac{1}{2\pi}\,\ln\rho \,+\, \OO(1)\,.
$$
The term with the minus sign is negative in $\,(l(l\!-\!1),l(l\!+\!1))\,$
provided $\,l\,$ is large enough. The other term is sign changing for
large $\,l\,$ so it has has a root. In view of Lemma~4.1 there is just
one $\,\mu_l\,$ such that
\begin{equation} \label{mu1}
\frac{2l\!+\!1}{l(l\!+\!1)-\mu_l}\,-\,\ln l\,+\,\OO(1)\,=\,0\;;
\end{equation}
this proves {\em (i)}. The relation (\ref{mu1}) yields also the
next two claims:
\begin{equation} \label{mu2}
\frac{2l\!+\!1}{l(l\!+\!1)-E}\,\leq\, \frac{2l\,+\,1}{l(l\,+\,1)-\mu_l}
\,=\,\ln l\,+\,\OO(1)
\end{equation}
and
\begin{equation}  \label{mu3}
\left|\, \frac{2l\!-\!1}{l(l\!-\!1)-E}\,\right|\,\leq\,
\frac{2l\!-\!1}{\mu_l-l(l\!-\!1)}\,=\,
\frac{2l\!-\!1}{2l\left(1-\,{1\!+\!\OO(1) \over \ln l}\right)}
\,=\,\OO(1)\,.
\end{equation}
Finally, {\em (iv)} follows from {\em (ii)} and {\em (iii)} which yield
$$
Z(\sqrt{\mu_l})\,=\, \frac{1}{4\pi}\,\left(\ln l+\OO(1)\right)\,-\,
\frac{1}{2\pi}\,\ln l\,+\,\OO(1)\,=\,-\,\frac{1}{4\pi}\,\ln l \,+\,\OO(1)
$$
and $\,\left|g(\sqrt{\mu_l})\right|=\,\frac{1}{4\pi}\,\ln l\,+\OO(1)\,$.
\quad \QED
\vspace{3mm}

Now we are in position to prove the main result of this subsection:
\begin{theorem}
Let $\,K_\eps:=\R\setminus\bigcup_{l=2}^\infty \left(\mu_l\!-\!
\eps(l)(\ln l)^{-2},\mu_l\!+\!\eps (l)(\ln l)^{-2}\right)\,$, where
$\,\eps(\cdot)\,$ is a positive strictly increasing function which
tends to $\,\infty\,$ and obeys the inequality $\,|\eps(x)|\leq
x\ln x\,$ for $\,x>1\,$. Then there is a positive $\,c\,$ such that
the transmission probability satisfies the bound
\begin{equation} \label{out reson}
|t(k)|^2 \le\, c\eps(l)^{-2}
\end{equation}
for $\,k^2\in K_\eps\cap (l(l\!-\!1),l(l\!+\!1))\,$ and any $\,l\,$
large enough. On the other hand, there are sharp resonance peaks
localized on $\,K_\eps\,$,
\begin{equation} \label{in reson}
|t(\sqrt{\mu_l})|^2 =\, 1+\OO\left((\ln l)^{-1} \right)
\end{equation}
as $\,l\to\infty\,$.
\end{theorem}
{\em Proof:} In the first part we are going to estimate the modulus
of the numerator in (\ref{t_bubble}) from below. The leading term at
high energies is the one with $\,\Delta(k)\,$; we shall estimate it on
the set $\,K_\eps\,$; \ie, away of the zeros of the coefficient.
Consider the neighborhood
$$
I_l\,:=\,\left(\mu_l\!-\!2l(\ln l)^{-1},
\mu_l\!+\!2l(\ln l)^{-1}\right)
$$
of $\,\mu_l\,$ on which the following estimates are valid:
\begin{eqnarray*}
\lefteqn{
{1\over 2k}\left(Z'(k)+(-1)^lg'(k)\right)\,=\,
\sum_{m=0}^{\infty}\, \frac{2m\!+\!1}{m(m\!+\!1)-E}
\left(1\!+\!(-1)^{l+m}\right) }
\\ &&\\ &&
=\,\left(\,\sum_{m=0}^{l-2}\,+\, \sum_{m=l+1}^{\infty}\,\right)
\frac{2m\!+\!1}{(m(m\!+\!1)-E)^2}(1\!+\!(-1)^{l+m})\,+\,
\frac{2(2l\!+\!1)}{(l(l\!+\!1)-E)^2}
\\ && \\ &&
\ge\,\frac{2(2l\!+\!1)}{(l(l\!+\!1)-E)^2}
\,\ge\,\frac{2(2l\!+\!1)}{\left(l(l\!+\!1)
-\mu_l+2l(\ln l)^{-1}\right)^2}
\\ && \\ &&
=\,\frac{2(2l\!+\!1)}{\left(2l(\ln l)^{-1}(2\!+\!\OO(1)\right)^2}
\,\ge\,c_1\,\frac{(\ln l)^2}{l}
\end{eqnarray*}
for some $\,c_1>0\,$. Combining this result with Lemma~4.1 we are able
to say how are the factors constituting $\,\Delta(k)\,$ separated from
zero for $\,E\in(l(l\!-\!1),l(l\!+\!1))\,$ which does not belong to the
interval $\,\left(\mu_l\!-\!\eps(l)(\ln l)^{-2}, (\mu_l\!+\!\eps(l)
(\ln l)^{-2}\right)\subset I_l\,$:
\begin{equation} \label{lower Delta}
\left|Z(k)+(-1)^{l}g(k)\right| \,\geq\, \eps(l)(\ln l)^{-2}c_1 l^{-1}
(\ln l)^2\,=\, c_1 l^{-1}\eps(l)\,.
\end{equation}
Now we begin estimating the modulus of the transmission amplitude
from below. In the following, $\,c_j\,$ always means a positive
constant. Using the expression (\ref{t_bubble}) we get a simple
lower bound,
\begin{equation} \label{upper t}
|t(k)|\,\leq\,{4\pi\rho \over\left|\,{1\!-\!2\pi Z(k) \over k|g(k)|}
\,-\,{\pi^2\Delta(k) \over |g(k)|} \left(k^{-1}\!-\!4k\rho^2 \right)\,
\right|}\,,
\end{equation}
is obtained by neglecting the imaginary part of the denominator. First
we shall show that
\begin{equation}\label{denom 1}
\frac{1-2\pi Z(E)}{k|g(k)|}\,=\,\OO(1)
\end{equation}
as $\,l\to\infty\,$. Using Lemmas~4.2 and 4.3 we estimate this
expression by
$$
{\left|\,\ln l+\OO(1)\right| \over kK}\,+\, {1\over 2\pi k}\,
\left|\, {\frac{2l\!+\!1}{l(l\!+\!1)-E} \,+\,
\frac{2l\!-\!1}{l(l\!-\!1)-E} \over
\frac{2l\!+\!1}{l(l\!+\!1)-E} \,-\,
\frac{2l\!-\!1}{l(l\!-\!1)-E}\,-2 }\,\right|\,\le\, c_2\,,
$$
where in the second term we have used the explicit lower bound on
$\,|g(k)|\,$, estimated the modulus of the fraction by
$\,{2l\!+\!1\over l\!-\!1}\,\le 5\,$, and employed finally
$\,k=\sqrt E=l+\OO(1)\,$. For the second term in the denominator of
(\ref{upper t}) we shall show that
\begin{equation}\label{denom 2}
\frac{\pi^2\Delta(k)}{|g(k)|}\,\left(k^{-1}\!-4k\rho^2\right)
\,\geq\, c_3\eps(l)
\end{equation}
holds for all $\,E\in(l(l\!-\!1),l(l\!+\!1)) \setminus (\mu_l\!-\!
\eps(l)(\ln l)^{-2},\mu_l\!+\!\eps(l)(\ln l)^{-2})\,$, which will give
(\ref{out reson}) with $\,c:= \left(c_3\over 4\pi\rho \right)^2$.
Consider first  the case $\,E\leq\mu_l\,$, when (\ref{lower Delta})
together with Lemmas~4.4(ii) yield
\begin{eqnarray*}
\lefteqn{\pi^2 \left|k^{-1}\!-4k\rho^2\right|
{ \left| Z(k)\!+\!(-1)^lg(k)\right|
\left| Z(k)\!-\!(-1)^lg(k)\right|\over
\left|\, \frac{1}{4\pi}\left(\,\frac{2l\!+\!1}{l(l\!+\!1)-E}
\,-\,\frac{2l\!-\!1}{l(l\!-\!1)-E}\right)\,+\, \OO(1)\,\right|} }
\\ && \\ &&
\ge\, 4\pi^2\rho^2 k \left|1\!-(2\rho k)^{-2}\right|
{ c_1 l^{-1} \eps(l) {1\over 2\pi} \left|\,
\frac{2l\!-\!1}{l(l\!-\!1)-E}\,-\, \ln l +\OO(1)\, \right|
\over \left| -\,{\,1\over 4\pi}\,\frac{2l\!-\!1}{l(l\!-\!1)-E}
\,-\, {1\over 4\pi}\,\ln l+\OO(1)\,\right| }
\\ && \\ &&
\ge \, c_4k\, {\eps(l)\over l}\,\ge\,c_5 \eps(l)\,,
\end{eqnarray*}
where in the last step we used again $\,k=l+\OO(1)\,$. Let further
$\,E>\mu_l\,$. We divide the argument into two parts. First we suppose
$$
\frac{2l\!+\!1}{l(l\!+\!1)-E}\,\leq\, 2\ln l\;;
$$
then
\begin{eqnarray*}
\lefteqn{\pi^2 \left|k^{-1}\!-4k\rho^2\right|
{ \left| Z(k)\!+\!(-1)^lg(k)\right|
\left| Z(k)\!-\!(-1)^lg(k)\right|\over
\left|\,\frac{1}{4\pi}\left(\,\frac{2l\!+\!1}{l(l\!+\!1)-E}
\,-\,\frac{2l\!-\!1}{l(l\!-\!1)-E}\right)\,+\, \OO(1)\,\right|} }
\\ && \\ &&
\ge\, 4\pi^2\rho^2 k \left|1\!-(2\rho k)^{-2}\right|
{ c_1 l^{-1} \eps(l) \left|\,\, \ln l +\OO(1)\, \right|
\over \left| \,\ln l+\OO(1)\,\right| }\,\ge\,c_6 \eps(l)
\end{eqnarray*}
by (\ref{lower Delta}) and Lemma~4.4(iii). On the other hand, if
\begin{equation} \label{case 2}
\frac{2l\!+\!1}{l(l\!+\!1)-E}\,\geq\,2 \ln l\,,
\end{equation}
the same expression is bounded from below by
\begin{eqnarray*}
\lefteqn{ 4\pi^2\rho^2 k \left|1\!-(2\rho k)^{-2}\right|
{ \left|\, {\,1\over 4\pi}\,\frac{2l\!+\!1}{l(l\!+\!1)-E}
\,+\OO(1)\,\right|\,
\left|\,\, {1\over 2\pi}\,\ln l +\OO(1) \,\right|
\over \left|\, {\,1\over 4\pi}\,\frac{2l\!+\!1}{l(l\!+\!1)-E}
\,+\OO(1)\,\right|} }
\\ && \\ &&
\,\ge\,  c_7k\,\left|1\!-(2\rho k)^{-2}\right|\, \ln l\,
\ge\, c_8\eps(l)\,, \phantom{AAAAAAAAAAAAA}
\end{eqnarray*}
where the denominator and the first term in the numerator have been
estimated by means of (\ref{mu1}) and (\ref{case 2}), in the second
term we have neglected one of the two terms of the  same sign, and
the last inequality follows from the fact that $\,x\ln x\geq\eps(x)\,$
by assumption. To conclude the proof of (\ref{out reson}), it is
sufficient to put $\,c_3:= \min\{c_5,c_6,c_8\}\,$ in (\ref{denom 2}).

The rest is easier; the existence of resonance peaks at which the
sphere is almost transparent follows from the relations
\begin{eqnarray*}
\left| t(\mu_l)\right| &\!=\!&
\left|\, {1\over 4\pi\rho \sqrt{\mu_l} g(\sqrt{\mu_l})}\,-\,
{Z(\sqrt{\mu_l}) \over g(\sqrt{\mu_l})}\,
\left(i+\, {1\over 2\rho \sqrt{\mu_l}} \right) \,\right|^{-1} \\ \\
&\!=\!& \left|\, \left(i+\, {1\over 2\rho \sqrt{\mu_l}} \right)
\left( 1+\OO((\ln l)^{-1}) \right)\,+\,
{1\over \rho \sqrt{\mu_l}} \left(\ln l+\OO(1)\right)\,\right|^{-1}
\\ \\
&\!=\!& \left|\, i+\OO((\ln l)^{-1})  \,\right|^{-1}
\,=\,1+\OO((\ln l)^{-1})\,,
\end{eqnarray*}
where we have used Lemma~4.4(iv). \quad \QED


\subsection{A ``bubble" array}

Having proved this we illustrate the behavior by numerical results. We find
interesting the behavior of $|t(k)|^2$. The curve oscillates almost regularly,
but with the amplitude spaning from small values to unity for sufficiently large
$k$. The lower enveloping curve of $|t(k)|^2$ behaves as $(E \ln E)^{-1}$ as we
proved above. However, the behavior of $|t(k)|^2$ resembles that of
$\delta'$--interaction, if an smoothed (locally averaged) curve is compared, as
already mentioned above. This suspicion can be supported by Fig. 7; it compares
an averaged $|t(k)|^2$ of a ``bubble" on the line and $|t(k)|^2$ of
$\delta'$--interaction. The averaging is done over ten neighboring peaks of a
given point. The strength of the $\delta'$--interaction is chosen so as to reach
the same value of $|t(k)|^2$ at a distant $k$. These two curves seem to have the
same ``asymptotic" behavior.

Fig. 8 shows the band spectrum of an infinite array of bubbles on the line.
What is different from analogical spectra of loop or comb arrays is the
concentration of bands to small values $\rho$ (note the logarithmic scale).


\renewcommand{\thesection}{\Alph{section}}
\renewcommand{\theequation}{\Alph{section}.\arabic{equation}}
\setcounter{section}{0}


\setcounter{equation}{0}
\section{A direct derivation of the comb--graph S--matrix}

To illustrate the advantages of the formulas (\ref{t_N}) and
(\ref{r_N}), we present two other ways to derive the comb--shaped
graph S--matrix.

{\em The first proof:} By brute force, looking directly for the
generalized eigenvector of $\,H_N\,$ at the energy $\,k^2\,$. One
takes for the line component of the wavefunction the following
Ansatz,
\begin{equation} \label{Ansatz}
f(x) \,=\, \left\{\begin{array}{lll}
e^{ikx} + r_N e^{-ikx} & \quad \dots & \quad x<0 \\ & \\
e_j e^{ikx} + f_j e^{-ikx} & \quad \dots & \quad L(j\!-\!1)<x<Lj \\ &
\\ t_N e^{ikx} & \quad \dots & \quad x>L(N\!-\!1)
\end{array}\right.
\end{equation}
for $\,j = 1,\dots,N-1\,$, while on the appendices we have
multiples of the above specified solution to (\ref{tooth}), $\,u_s=
\beta_s u_\ell\,, \; s=1,\dots,N\,$. We have to find the coefficients
for which such a vector belongs {\em locally} to the domain of the
Hamiltonian.

Substituting from (\ref{Ansatz}) to (\ref{comb bc}) we get a system
of linear equations; denoting
\begin{equation} \label{notation1}
e_0 := 1,\quad f_n := 0,\quad e_N := t_N,\quad f_0 := r_N,
\end{equation}
we can write it concisely as
\begin{eqnarray}
e_j\eps^j +f_j\eps^{-j} -e_{j+1}\eps^j -f_{j+1}\eps^{-j} &\!=\!& 0\,,
\label{eq1} \\
e_{j+1}b\eps^j + f_{j+1}b\eps^{-j} + \beta_{j+1}(cu'_\ell(0)-u_\ell(0))
&\!=\!& 0\,, \label{eq2} \\
(ik\!-\!d)\eps^je_{j+1} + (ik\!+\!d)\eps^{-j}f_{j+1} + ik\eps^je_j +
ik\eps^{-j}f_j + \beta_{j+1}bu_\ell'(0) &\!=\!& 0 \label{eq3}
\end{eqnarray}
for $\,j = 0,\dots,N\!-\!1\,$. First we express $\,e_j\,$ from
(\ref{eq1}): by induction we check the relation
\begin{equation} \label{e_j}
e_j \,=\, 1 + f_0 + \left(\sum_{l=1}^{j-1}f_\ell\eps^{-2(l-1)}\right)
(\eps^{-2}\!-\!1) - f_j\eps^{-2(j-1)}.
\end{equation}
Using (\ref{eq2}) we exclude $\,e_{j+1}\,$ from (\ref{eq3}), which
yields
\begin{equation}\label{ident1}
(2ik\!-\!d)\eps^{-j}f_j - de_j\eps^j - 2ik\eps^{-j}f_{j+1} +
\beta_{j+1}bu_\ell'(0) \,=\, 0.
\end{equation}
Next we exclude $\,\beta_j\,$ from (\ref{eq2}) and (\ref{ident1});
substituting then for $\,e_j\,$ from (\ref{e_j}) we get
\begin{eqnarray}\label{system2}
-f_1G + f_0[G\!-\!F] &\!=\!& F\,,\nonumber \\
-G\eps^{-j}f_{j+1} + \left\lbrack (\eps^2\!-\!1)+G \right\rbrack
\eps^{-j}f_j - \eps^jF(1\!-\!\eps^2)\left(\sum_{l=1}^{j-1}
f_\ell\eps^{-2l}\right) - \eps^jf_0 &\!=\!& 0\,, \phantom{AAA}
\end{eqnarray}
where $\,j = 1,\dots,N\!-\!1\,$ and $\,F,G\,$ are the quantities
defined above.  The last system can be transformed in such a way that
its matrix is tridiagonal; to this end we add the $\,j$--th equation
to the $\,(j\!-\!1)$--th one multiplied by $\,\eps\,$. The resulting
system has the following form,
\begin{eqnarray}\label{system3}
-f_1d(cu_\ell'-u_\ell)(0) + f_0[G\!-\!F] &\!=\!& F\,, \nonumber \\
-Gf_{j+1} + [(1+\eps^2)G+(\eps^2\!-\!1)F]f_j - G\eps^2f_{j-1} &\!=\!&
0\;,
\end{eqnarray}
where $\,j = 2,\dots,N\!-\!1\,$. In terms of $\,\beta,\,z\,$ this can
be rewritten as
\begin{eqnarray}\label{system4}
f_0(1+i\beta) - f_1 &\!=\!& -i\beta\,, \nonumber \\
\eps^{-1}f_{j+1} - 2zf_j + \eps f_{j-1} &\!=\!& 0\,,\quad j =
1,\dots,N\!-\!2\,, \\
-2zf_{N\!-\!1} + \eps f_{N-2} &\!=\!& 0\;; \nonumber
\end{eqnarray}
we have taken into account here that $\,f_N=0\,$. Now it is
straightforward to find $\,f_0=r_N\,$ as a ratio of the corresponding
determinants, $\,r_N= D^{(2)}_N/D^{(1)}_N\,$, with
$$
D_N^1 = -i\beta D_{N-1}\,, \qquad D_N^2 =(1\!+\!i\beta)D_{N-1}\! +
\eps D_{N-2}\,,
$$
where
$$
D_N\,:=\, \left|\begin{array}{ccccc}
     -2z & \eps^{-1} & \ldots & 0 & 0\\
     \eps & -2z & \eps^{-1} & \ldots & 0\\
     \vdots & \vdots & \vdots & \vdots & \vdots \\
     0 &\ldots & \eps & -2z & \eps^{-1}\\
     0 & 0 & \ldots & \eps & -2z
     \end{array}\right|
$$
and $\,D_0:=0\,$. The above determinants satisfy the recursive
relation $\,D_{N+2} + 2zD_{N+1} + D_N = 0\,$, so by induction we get
$\,D_N= (-1)^NU_N(z)\,$, \ie, the first formula in (\ref{comb rt_N}).

In a similar way we find the transmission amplitude. We start again
from (\ref{eq1}) from which we express $\,f_j\,$,
$$
f_j = \sum_{l=j}^{N-1}\eps^{2l}(e_{l+1}\!-e_\ell),\qquad
j=0,\dots,N\!-\!1\,.
$$
Transforming the system we arrive at an analogy of (\ref{system4})
in the variables $\,e_j\,$,
\begin{eqnarray*}
-2ze_1 + \eps e_2 &\!=\!& -\eps^{-1}\,, \\
\eps e_{j+2} - 2zef_{j+1} + \eps^{-1}e_j &\!=\!& 0\,,\qquad
j=1,\dots,N\!-\!2\,, \\
e_N(1+i\beta) - e_{N-1} &\!=\!& 0\,,
\end{eqnarray*}
which has to be solved for $\,e_N=t_N\,$. A disadvantage of this
argument is that it gives a little insight into the way in which
graph appendices contribute to the S--matrix.
\vspace{2mm}

{\em The second proof:} By S--matrix factorization described in
Section 2.2. The on--shell  S--matrix of a single tooth attached at
$\,x=\xi\,$ is
\begin{equation} \label{singleS}
S = \left(\begin{array}{cc}
e^{2ik\xi}r & t\\
t & e^{-2ik\xi}r
\end{array}\right)\,,
\end{equation}
where $\,t:=t_1\,$ and $\,r:=r_1\,$ are the corresponding amplitudes
of a tooth attached at $\,x=0\,$ which can be adopted from
\cite{ESe}. The matrix (\ref{singleS}) is unitary since $\,t/r\,$ is
purely imaginary. The mentioned factorization technique yields for
the amplitudes $\,r_N\,$ and $\,t_N\,$ the following recursive
relations
\begin{equation}\label{recur1}
r_{N+1} = r_N + \frac{\eps^{2N}rt_N^2}{1\!-\!\eps^2rr_N}\,,\qquad
t_{N+1} = \frac{tt_N}{1\!-\!\eps^2rr_N}\,,
\end{equation}
which are equivalent to
\begin{equation} \label{recur2}
t(r_{N+1}\!-r_N) = \eps^{2N}rt_{N+1}t_N\,,\qquad
(1\!-\!\eps^2rr_N)t_{N+1} = tt_N\,.
\end{equation}
Substituting from (\ref{comb rt_N}) into (\ref{recur2}) and employing
the relations between Chebyshev polynomials we check that $\,r_N\,$ and
$\,t_N\,$ satisfy indeed (\ref{recur1}). Certain disadvantage of this
argument is that the closed form of the amplitudes (\ref{comb rt_N})
has to be guessed or derived by other means.


\subsection*{Acknowledgment}

The research has been partially supported by GA AS under the
contract $\#$ 1048801.

\section*{Figure captions}

   \begin{description}
   \item{\em Figures 2, 5, 8\quad} available upon request from
   {\em tater@ujf.cas.cz}
   \item{\bf Figure 1\quad} A loop--graph scatterer in a magnetic field
   \item{\bf Figure 2\quad} The relation of transmission probabilities of
   a finite array of loops ($N = 6$) and the spectrum of the corresponding
   infinite system, $L_1 = 0.5$, $L_2 = 1.5$, $\ell = \sqrt{2}$,
   $\alpha_1 = -1$, $\alpha_2 = -2$.
   \item{\bf Figure 3\quad} A comb-shaped graph
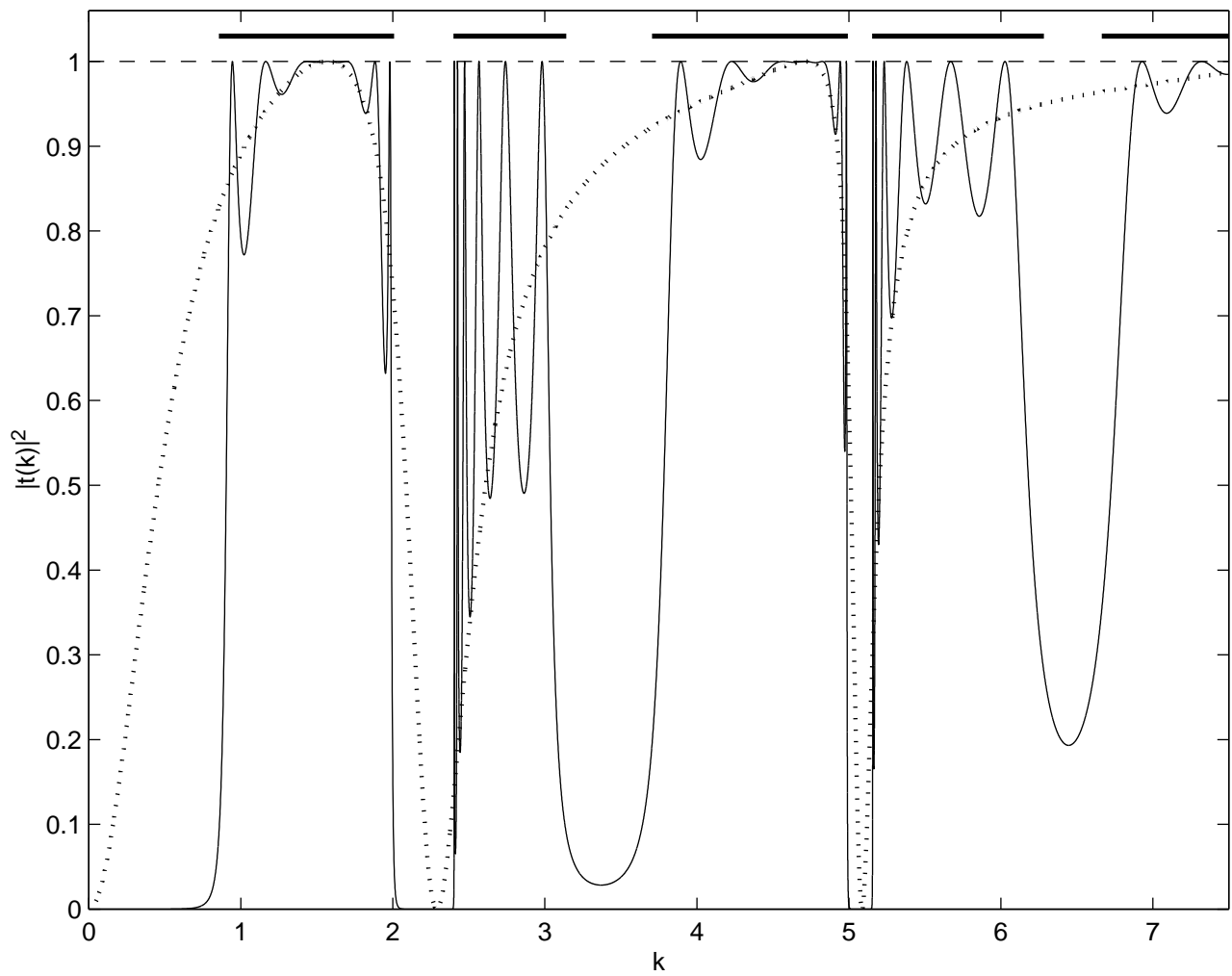
   \item{\bf Figure 4\quad} The relation of transmission probabilities of a
   single appendix (dotted line),
   a finite array of appendices ($N = 7$) (full line), and the spectrum of the
   corresponding infinite system (thick lines above). The parameters are $b = 1.2$,
   $c = 0.5$, $d = 0$, $L = 1$, and $\ell = 1$.
   \item{\bf Figure 5\quad} The band spectra of infinite arrays of appendices.
   The upper figure is for $b = 1$, $c = d = 0$, and the lower one for $b = 3$,
   $c = 0.5$, and $d=-11$. The spacing is chosen $\ell = 1$ in both cases. The
   full lines show the role of perfect transmission (the lines in the bands) and
   total reflection (the lines in the gaps) for a single appendix.
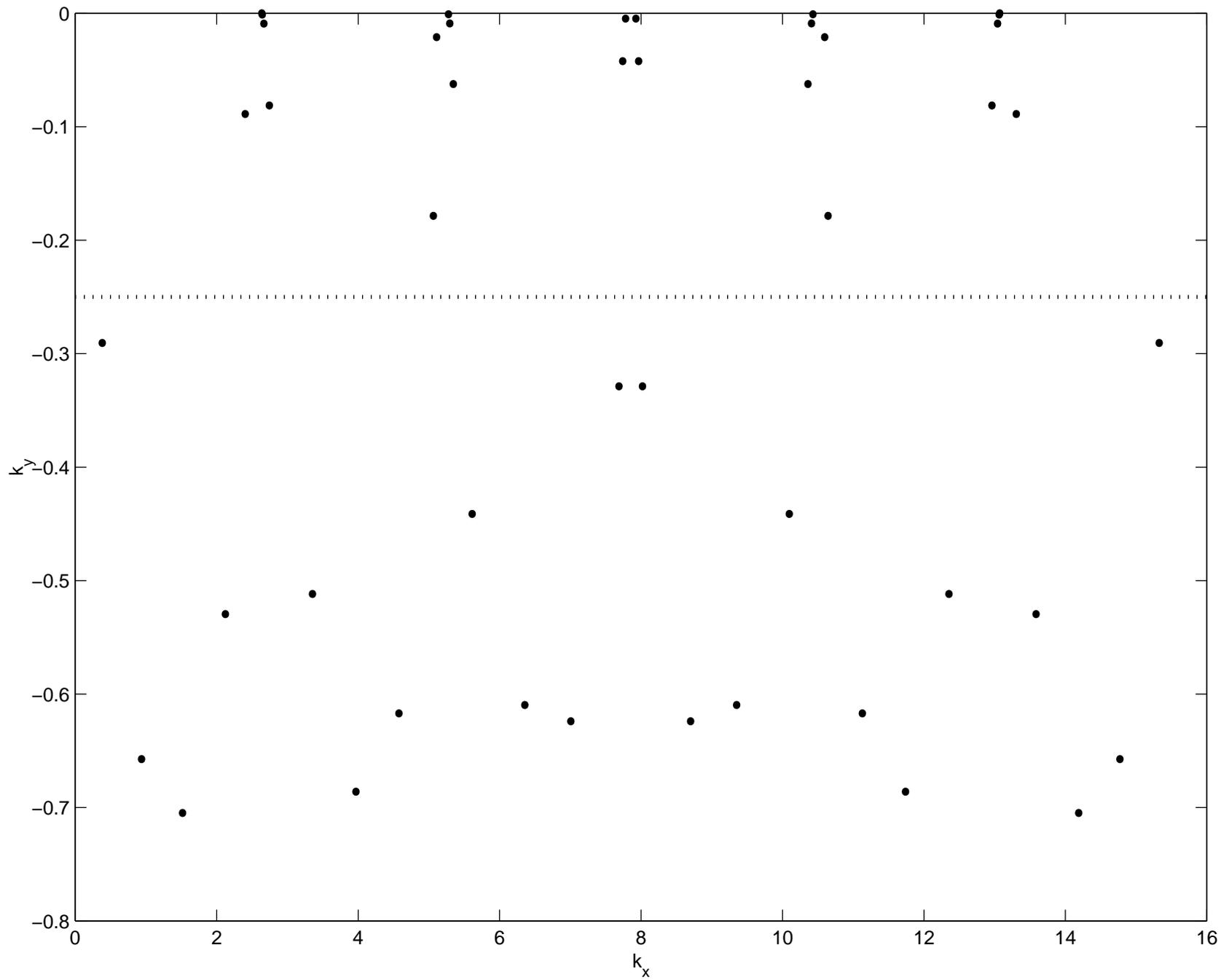
   \item{\bf Figure 6\quad} This figure shows positions of resonances for $N = 5$
   appendices and $b = 0.4$, $c = d = 0$, and $L = 1.2$. The dotted line separates
   the two sets of resonances. The resonances closer to the real axis have their
   origin in $N$-fold eigenvalue for $b = 0$; increasing $|b|$ lifts the
   degeneracy and pushes the resonances from the real line. The rest of the
   resonances (below the dotted line) has the origin in the serial arrangement of
   individual scatterers.
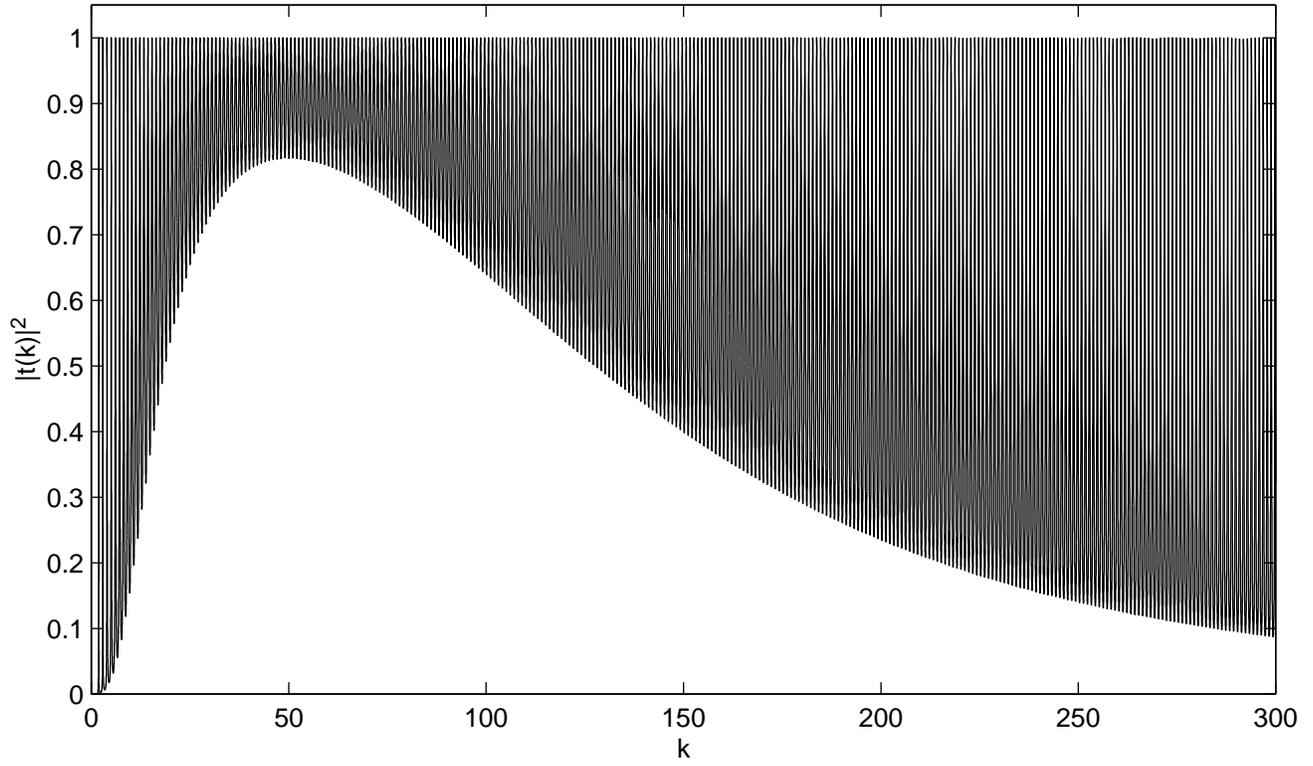
   \item{\bf Figure 7\quad} The transition probability of a sphere of unit
   radius on a line ($\rho = 0.01$). The lower graph compares the asymptotic
   behavior of $|t(k)|^2$ for $\delta'$--interaction with suitably chosen strength
   and that of averaged transmission probability plotted in the upper graph.
   The averaging is done over ten neighboring peaks of each point.
   \item{\bf Figure 8\quad} Band spectrum of an infinite ``bubble" array. The
   spheres are of unit radius, the spacing is $\ell = 1$ (upper figure) and
   $\ell = 0.01$ (lower figure), $\rho$ is the contact radius.

   \end{description}

\end{document}